\documentclass[12pt]{article}
\input epsf.sty
\def\ZZZ{{\hbox{ Z\kern-1.6mm Z}}}

\newcommand{\ra}{\rangle}
\newcommand{\la}{\langle}

\newcommand{\vp}{\varphi}

\newcommand{\tl}{\lambda}

\newcommand{\BB}{{\cal B}}

\newcommand{\OO}{{\cal O}}

\newcommand{\PP}{{\cal P}}

\newcommand{\wt}{\widetilde}
\newcommand{\wh}{\widehat}
\newcommand{\wc}{\check}

\newcommand{\RR}{{\cal R}}
\newcommand{\NN}{{\cal N}}
\newcommand{\TT}{{\cal T}}

\newcommand{\omk}{\omega_n(\vec k)}
\newcommand{\onk}{\omega^{(N)}_{\vec k_\perp}}

\newcommand{\be}{\begin{equation}}
\newcommand{\ee}{\end{equation}}
\newcommand{\ben}{\begin{eqnarray}\displaystyle}
\newcommand{\een}{\end{eqnarray}}
\newcommand{\refb}[1]{(\ref{#1})}
\newcommand{\p}{\partial}
\newcommand{\sectiono}[1]{\section{#1}\setcounter{equation}{0}}

\def\one{{\hbox{ 1\kern-.8mm l}}}
\def\zero{{\hbox{ 0\kern-1.5mm 0}}}
 
\begin{document}
{}~
{}~
\hfill\vbox{\hbox{hep-th/0402157}
}\break
 
\vskip .6cm
\begin{center}
{\Large \bf 
Rolling Tachyon Boundary State, Conserved Charges

\medskip

and Two Dimensional String Theory
}

\end{center}

\vskip .6cm
\medskip

\vspace*{4.0ex}
 
\centerline{\large \rm
Ashoke Sen}
 
\vspace*{4.0ex}

\centerline{\large \it Harish-Chandra Research Institute}

\centerline{\large \it  Chhatnag Road, Jhusi,
Allahabad 211019, INDIA}
 
\centerline{E-mail: ashoke.sen@cern.ch,
sen@mri.ernet.in}
 
\vspace*{5.0ex}
 
\centerline{\bf Abstract} \bigskip

The boundary state associated with the rolling tachyon solution on an
unstable D-brane contains a part that decays exponentially in the
asymptotic past and the asymptotic future, but it also contains other
parts which either remain constant or grow exponentially in the past or
future.  We argue that the time dependence of the latter parts is
completely determined by the requirement of BRST invariance of the
boundary state, and hence they contain information about certain conserved
charges in the system. We also examine this in the context of the unstable
D0-brane in two dimensional string theory where these conseved charges
produce closed string background associated with the discrete states, and
show that these charges are in one to one correspondence with the symmetry
generators in the matrix model description of this theory.

\vfill \eject
 
\baselineskip=18pt

\tableofcontents

\sectiono{Introduction and Summary} \label{sintro}

Unstable D-branes in bosonic and superstring theory admit time dependent 
solutions describing the rolling of the tachyon away from the maximum of 
the potential\cite{0202210,0203211,0203265}. At the open string tree level 
these solutions are described by solvable boundary conformal field 
theories. Using this description one can calculate the sources for various 
closed string states produced by this rolling tachyon solution. The 
information about these sources can be summarized in the boundary state 
associated to the solution, which is a  ghost number 3 
state in the Hilbert space of closed string states.

The boundary state associated with the rolling tachyon solution
can be divided into two parts. 
The first part gives rise to sources for various closed string states 
which fall off exponentially both in the asymptotic future and 
in the asymptotic past. Thus this part induces a closed string background 
that satisfy source free closed string field equations in the asymptotic 
past and the asymptotic future. Computation of total energy stored in the 
closed string field shows that while the amount of energy stored in a 
given mode of a fixed mass is small compared to the total energy of the 
D-brane in the weak coupling limit, the total amount of energy stored in 
the closed string field becomes infinite when we sum over all the 
modes\cite{0303139,0304192,0402124}.
Naively, this suggests that the tree level open string description of the 
system breaks down due to the backreaction of the closed string emission. 
However, a different viewpoint, proposed in 
\cite{0305011,0306137,0308068}, is that the closed strings do not 
invalidate the open string results, but simply provide 
a dual 
description of the same results. From this viewpoint, 
quantum open string theory provides a complete description of the unstable 
D-brane system, and there is no need to include the effect of closed 
string emission in the open string analysis. Ehrenfest theorem will then 
tell us that in the weak coupling limit tree level open string theory 
provides a 
complete understanding of the dynamics of the system. Evidence for this 
conjecture comes from the observation that many of the properties of the 
final system predicted by the tree level open string analysis, {\it e.g.} 
vanishing 
pressure and dilaton charge, agree with the properties of the final closed 
string field configuration in the weak coupling limit.

This conjecture can be put on a firm footing in two dimensional string 
theory based on the conformal field theory of a time like scalar field 
with central charge 1, and the Liouville field theory of central charge 
25 \cite{KPZ,DAVID,DISKAW}.\footnote{This has also been generalized to two 
dimensional string 
theories with world-sheet supersymmetry\cite{0307083,0307195}, but 
for simplicity we 
shall focus our attention on two dimensional bosonic string theory only.} 
This theory admits an unstable D0-brane and rolling tachyon 
solutions on this D0-brane.
On the other hand this string theory has 
a dual description as a matrix model\cite{GROMIL,BKZ,GINZIN}, which, in 
turn is described by a theory of free fermions in an inverted harmonic 
oscillator potential. The vacuum of this theory is described by a state 
where all states below a given level (fermi level) are filled, and all 
states above this 
level are empty. By expressing the closed string state produced by the 
rolling tachyon configuration in
the language of 
this free fermion field theory one finds that this represents a state 
where a single fermion is excited from the fermi level to some energy 
above the fermi 
level\cite{0304224,0305159,0305194,0307083,0307195,0307221,0308047,0310106,
0312135,0312163,0312192,0310287,0312196,0401067}. Thus the dynamics of a 
D0-brane 
is described completely by the theory of a single particle moving in an 
inverted harmonic oscillator potential, 
with the additional constraint that the energy levels of the 
system below the fermi level are removed by hand. This can be 
regarded as the exact open string description of the system.
The closed string 
fields in this theory are obtained by bosonizing the free field theory of 
fermions in the inverted harmonic oscillator 
potential\cite{DASJEV,SENWAD,GROSSKLEB}. This clearly demonstrates that 
the closed strings provide a description that is dual to the open string 
description, but the open string theory is capable of providing a complete 
description of 
the system.

All this analysis has been done with only one part of the boundary 
state associated with the
rolling tachyon solution. But both in the critical string theory and in 
the two dimensional string theory the boundary state
has another part which gives rise to sources which do not vanish in the 
asymptotic
past of future, but either remain constant\cite{0203265} or grow 
exponentially\cite{0208196} either
in the past or in the future. Naively, this would again indicate that
these exponentially growing closed string fields invalidate the classical
open string description of the system. However the open-closed string 
duality
conjecture stated above would lead one to believe that this is not so.
What this may be indicating is an inadequacy in the closed string
description rather than in the open string description. As an analogy 
we 
can cite the example of closed string field configurations produced by 
static stable D-branes. Often the field configuration hits a singularity 
near the core of the brane. However we do not take this as an indication 
of the breakdown of the open string description. Instead it is a 
reflection of the inadequacy of the closed string description. 

It still makes sense however to explore whether the exponentially growing 
source 
terms contain any physical information about the system. In this paper we 
argue that they contain information about some conserved charges. The 
argument relies on the fact 
that while the requirement of BRST invariance does not put any constraint 
on the first part of the boundary state, it fixes the time dependence of 
the various source terms coming from the second part. 
While in the critical string theory we do not have any 
independent way of 
verifying these conservation laws, in two dimensional string theory we can 
find additional 
support for this interpretation by identifying these conserved charges in 
the matrix model description.

The rest of the paper is organised as follows. In section \ref{sboundary} 
we split the boundary state of the critical string theory into two 
parts each of which is separately BRST invariant. The first part 
vanishes in the asymptotic past and asymptotic future, but the 
second part contains constant as well as exponentially growing 
terms. We show that whereas the time dependence of the first part 
is not fixed by the requirement of BRST invariance of the boundary 
state, the time dependence of the second part does get fixed by 
this requirement. This leads to the suggestion that this part 
contains information about conserved charges of this 
system\cite{0203265,0309074}. 
These conserved charges are shown to be labeled by SU(2) quantum numbers 
$(j,m)$ with the
restriction $-(j-1)\le m\le (j-1)$. We also find the dependence of these 
charges on the parameter characterizing the rolling tachyon solution.

In section \ref{sclosed} we analyze the closed string field configuration 
produced by different parts of the boundary state. Since the first part 
vanishes in the limit $x^0\to\pm\infty$, this produces source free 
on-shell closed string background in these 
limits\cite{0303139,0304192,0402124}. 
For the second part the source terms do not vanish in the 
$x^0\to\pm\infty$ limit. However, since the sources are 
localized 
at the location of the original D-brane, we get source free on-shell 
closed string background away from the location of the brane.

In section \ref{s2d} we repeat the analysis of section \ref{sboundary} for 
the D0-brane of the two dimensional string theory. We split the boundary 
state associated with the rolling tachyon solution on the D0-brane into 
two parts each of which is
separately BRST invariant, and following arguments identical to that in 
the 
case of critical string theory we show that while the first part vanishes 
in the asymptotic past and future, the second term contains information 
about conserved charges in the system. In section \ref{s2closed} we 
analyze the closed string fields produced by this boundary state. The 
first part produces on-shell closed string tachyon field in the asymptotic 
past and 
future\cite{0305159}, whereas the second part produces on-shell closed 
string 
background which are analytic continuation of the discrete 
states\cite{LIAN,9108004,9201056} in the 
euclidean two dimensional string theory.

Finally in section \ref{smatrix} we identify 
these 
conserved charges in the matrix model description of the two dimensional 
string theory. We begin this section with a review of the description of 
the D0-brane in the matrix model.
We then explicitly identify a set of conserved charges in the 
matrix 
model following 
\cite{SENWAD,MOORESEI,UTTG-16-91,9108004,9110021,9201056,9209036,9302106,
9507041}, 
and 
find the precise relation between these conserved chrges and those in the 
continuum description by comparing their dependences on the time 
coordinate $x^0$ and the parameter $\lambda$ labelling the rolling tachyon 
solution. 
In particular we establish a one to one correspondence between the 
conserved charges in the two descriptions following this line of argument. 

Each of the sections also contains a large amount of
material where we review the relevant aspects 
of the 
decaying part of the boundary state before turning our attention to 
the 
constant and the exponentially growing parts.

\sectiono{Boundary State for the Rolling Tachyon in Critical String 
Theory} \label{sboundary}

We begin with a D-$p$-brane in critical bosonic string theory in flat 
(25+1) dimensional space-time. The rolling 
tachyon solution, parametrized by the constant $\tl$, is obtained by 
deforming the conformal field theory describing the D-$p$-brane by a 
boundary term\cite{0203211,0203265}
\be \label{ehx3}
\tl \, \int dt \, \cosh(X^0(t))\, .
\ee
We are using $\alpha'=1$ unit.
Under a Wick rotation $X^0\to iX$, this becomes:
\be \label{ehx1}
\tl \, \int dt \, \cos(X(t))\, .
\ee
The boundary state associated with the Euclidean D-brane, corresponding to 
the deformation \refb{ehx1}, is given by:
\be \label{efg10}
|\BB\ra = \TT_p \, \,  |\BB\ra_{c=1} \otimes |\BB\ra_{c=25} \otimes
|\BB\ra_{ghost}\, ,
\ee
where $\TT_p$ is the tension of the D-$p$-brane, $|\BB\ra_{c=1}$ denotes 
the boundary state associated with the
$X$ direction, $|\BB\ra_{c=25}$ denotes the boundary state
associated with the other 25 directions $X^1, \ldots X^{25}$, and
$|\BB\ra_{ghost}$ denotes the
boundary state associated with the ghost direction. We have:
\be \label{efg11}
|\BB\ra_{c=25} =
\int {d^{25-p} k_\perp \over (2\pi)^{25-p}} \,
\exp\left(\sum_{s=1}^{25}\sum_{n=1}^\infty
{1\over
n} \, (-1)^{d_s} \, \alpha^s_{-n} \bar
\alpha^s_{-n} \right)
|k_\parallel=0,
k_\perp\ra\, .
\ee
and
\be \label{efg12}
|\BB\ra_{ghost} = \exp\left(-\sum_{n=1}^\infty (\bar b_{-n} c_{-n} +
b_{-n}
\bar c_{-n})
\right) (c_0+\bar c_0)c_1\bar c_1 |0\ra\, .
\ee
Here $d_s=1$ if $X^s$ has Neumann boundary condition and 0 if $X^s$ has 
Dirichlet boundary condition,
$\vec k_\parallel$ denotes spatial momentum
along the D-$p$-brane, $k_\perp$ denotes
momentum transverse to the D-$p$-brane, $\alpha^s_n$, $\bar 
\alpha^s_n$ denote the oscillators associated with the world-sheet scalar 
field $X^s$ and $b_n$, $\bar b_n$, $c_n$, $\bar c_n$ denote the ghost 
oscillators. For $\tl =0$, $|\BB\ra_{c=1}$ becomes the 
standard 
boundary state for $X$ with Neumann boundary condition:
\be \label{efg13}
|\BB\ra_{c=1}|_{\tl=0} = \exp\left(-\sum_{n=1}^\infty
{1\over
n} \alpha_{-n} \bar
\alpha_{-n} \right) |k=0\ra\, ,
\ee
where $k$ labels momentum along $X$. 

For non-zero $\tl$, it is convenient to express $|\BB\ra_{c=1}$
as a sum of two 
terms\cite{9402113,9811237,0203265,0208196,0212248,0305177,0308172}:
\be \label{ebs1}
|\BB\ra_{c=1} = \exp\left(\sum_{n=1}^\infty \, {1\over n}
\alpha_{-n}\bar\alpha_{-n}\right)\, f(X(0)) \, |0\ra +
|\wt\BB\ra_{c=1} \, ,
\ee
where
\ben \label{edeffagain}
f(x) &=& {1\over 1 + \sin(\pi\tl) e^{ix}} + {1\over 1 +
\sin(\pi\tl) e^{-ix}} - 1\, , \nonumber \\
&=& \sum_{n\in Z} (-1)^n \sin^{|n|}(\pi\tl) \, e^{inx} \, .
\een
Note the difference in sign in the exponents of \refb{efg13} and 
\refb{ebs1}. For any momentum $k$, $\exp\left(\sum_{n=1}^\infty \, {1\over 
n}
\alpha_{-n}\bar\alpha_{-n}\right)|k\ra$ is annihilated by 
$L^X_{n}-\bar L^X_{-n}$ where $L^X_n$ and $\bar L^X_n$ denote the Virasoro 
generators of the $c=1$ conformal field theory. Since $|\BB\ra_{c=1}$ must 
be annihilated by $L^X_{n}-\bar L^X_{-n}$, it follows that 
$|\wt\BB\ra_{c=1}$ must also be annihilated by $L^X_{n}-\bar L^X_{-n}$ and 
hence must be a linear combination of the Ishibashi states\cite{ishibashi} 
built upon various left-right symmetric primary states in the $c=1$ 
conformal 
field theory. These primaries are labelled by the SU(2) quantum numbers 
$(j,m)$ $(-j\le m\le j$, $j-m$ integer), with $2m$ denoting the 
$X$ 
momentum carried by the state, and 
$(j^2, j^2)$ being the conformal weight of the state. 
Thus the primary state $|j,m\ra$ has the form:
\be \label{epriform}
|j,m\ra = \wh \PP_{j,m} \, e^{2 \, i \, m \, X(0)} |0\ra\, ,
\ee
where $\wh \PP_{j,m}$ is some combination of the $X$ oscillators of level 
$(j^2 - m^2, j^2 - m^2)$.
We shall normalize $\wh \PP_{j,m}$ such that when we express 
$\exp\left(\sum_{n=1}^\infty \, {1\over n}
\alpha_{-n}\bar\alpha_{-n}\right)e^{2im X(0)}|0\ra$ as a linear 
combination 
of the Ishibashi states built on various primaries, the Ishibashi state 
$|j,m\ra\ra$ built on the primary 
$|j,m\ra$ appears with coefficient 1:
\be \label{edefjm}
\exp\left(\sum_{n=1}^\infty \, {1\over
n}
\alpha_{-n}\bar\alpha_{-n}\right)e^{2im X(0)}|0\ra = \sum_{j\ge |m|} 
|j,m\ra\ra\, .
\ee 
The 
complete contribution to $|\BB\ra_{c=1}$ from the Ishibashi states built 
over the primaries 
$|j,\pm j\ra$, as well as part of the contribution 
from 
the other Ishibashi states, are included in the first term on the right 
hand side of \refb{ebs1}. 
Thus the second term must be a linear combination of Ishibashi states 
built on $|j,m\ra$ with $m\ne \pm j$:
\be \label{esecondform}
|\wt\BB\ra_{c=1} = \sum_{j} \, \sum_{m=-j+1}^{j-1} \, f_{j,m}(\tl) \, 
|j,m\ra\ra\, ,
\ee
where
$f_{j,m}(\tl)$ are some
functions of the 
parameter $\tl$. These are given by\cite{9402113,9811237}:
\be \label{efjm}
f_{j,m}(\tl) = D^j_{m,-m}(2\pi\tl) \, {(-1)^{2m} \over D^j_{m,-m}(\pi)} - 
(-1)^{2m} \, \sin^{2|m|}(\pi\tl)\, ,
\ee
where $D^j_{m,m'}(\theta)$ are the representation matrices of the SU(2) 
group element $e^{i\theta\sigma_1/2}$ in the
spin 
$j$
representation.
The second term in \refb{efjm} represents the effect of subtracting from 
$|\BB\ra_{c=1}$ the 
contribution due to the first term in \refb{ebs1}.
The fact that the total contribution to $|\BB\ra_{c=1}$ from $|j,m\ra\ra$ 
is
proportional to $D^j_{m,-m}(2\pi\tl)$ was 
shown in \cite{9402113,9811237}. The constant of proportionality is found 
by using the condition
\be \label{efjm2}
f_{j,m}\left({1\over 2}\right) = 0\, .
\ee
This relation arises
as follows. As 
$\tl\to {1\over 2}$, the system approaches an array of 
D-branes with Dirichlet boundary condition on $X$, situated at 
$x=(2k+1)\pi$ for integer $k$. 
On the other hand, from \refb{edeffagain} one can show that
\be \label{elimit}
\lim_{\tl\to{1\over 2}} \, f(x) = 2\pi \, \sum_{k\in Z} \, \delta\left(x - 
(2k+1) 
\, 
\pi\right)\, .
\ee
In this case the first term in 
\refb{ebs1} reproduces the complete contribution to the boundary 
state, and hence the second term must vanish. This is the reason why the 
functions $f_{j,m}(\tl)$ must vanish in the $\tl\to {1\over 2}$ limit. For 
later use we quote here the form of $f_{j,m}(\tl)$ for some specific 
$(j,m)$:
\be \label{especial}
f_{1,0}(\tl) = - 2 \cos^2(\pi\tl), \qquad f_{{3\over 2},{1\over 2}}(\tl) = 
3
\sin(\pi\tl)\cos^2(\pi\tl)\, ,
\ee
etc.

The boundary state in the Minkowski theory with boundary interaction 
\refb{ehx3} is then obtained by the replacement $X\to -i X^0$ in the 
Euclidean boundary state. If $|\wh\BB\ra_{c=1}$ denotes the continuation 
of $|\wt\BB\ra_{c=1}$ to Minkowski space, 
\be \label{ehbdef}
|\wh\BB\ra_{c=1} = |\wt\BB\ra_{c=1} \big|_{X\to - i X^0}\, ,
\ee
then the complete boundary state 
in the Minkowski
space is given by:
\be \label{ebs2}
|\BB\ra = |\BB_1\ra + |\BB_2\ra\, ,
\ee
where
\ben \label{ebs3}
|\BB_1\ra &=& \TT_p\,  \exp\left(-\sum_{n=1}^\infty \, {1\over n}
\alpha^0_{-n}\bar\alpha^0_{-n}\right)\, \wt f(X^0(0))\, |0\ra \nonumber \\
&& \otimes
\int {d^{25-p} k_\perp \over (2\pi)^{25-p}}\,
\exp\left(\sum_{n=1}^\infty
\sum_{s=1}^{25} (-1)^{d_s} \, {1\over
n} \,  \alpha^s_{-n} \bar
\alpha^s_{-n} \right) |\vec k_\parallel=0,
\vec k_\perp\ra\, \nonumber \\
&& \otimes
\exp\left(-\sum_{n=1}^\infty (\bar b_{-n} c_{-n} +
b_{-n}
\bar c_{-n})
\right) (c_0+\bar c_0)c_1\bar c_1 |0\ra\, ,
\een
and
\ben \label{ebs4}
|\BB_2\ra &=& \TT_p\, |\wh\BB\ra_{c=1} \, \otimes
\int {d^{25-p} k_\perp \over (2\pi)^{25-p}} \,
\exp\left(\sum_{n=1}^\infty
\sum_{s=1}^{25} (-1)^{d_s} \,{1\over
n} \,  \alpha^s_{-n} \bar
\alpha^s_{-n} \right) |\vec k_\parallel=0,
\vec k_\perp\ra\, \nonumber \\
&& \otimes
\exp\left(-\sum_{n=1}^\infty (\bar b_{-n} c_{-n} +
b_{-n}
\bar c_{-n})
\right) (c_0+\bar c_0)c_1\bar c_1 |0\ra\, ,
\een
where 
\be \label{efdef}
\wt f(x^0) = f( - i x^0) = {1\over 1 + \sin(\pi\tl) e^{x^0}} + {1\over 1 +
\sin(\pi\tl) e^{-x^0}} - 1\, .
\ee

Using the fact that $L^{X^0}_n-\bar L^{X^0}_{-n}$ annihilates 
$\exp\left(-\sum_{n=1}^\infty {1\over n}
\alpha^0_{-n}\bar\alpha^0_{-n}\right) |k^0\ra$
it is easy to verify that $|\BB_1\ra$ is BRST invariant, {\it i.e.}
\be \label{ebrinv}
(Q_B+\bar Q_B) |\BB_1\ra = 0\, .
\ee
Indeed we have the stronger relation
\ben \label{estrong}
&& (Q_B+\bar Q_B)\, \bigg[
\exp\left(-\sum_{n=1}^\infty {1\over n}
\alpha^0_{-n}\bar\alpha^0_{-n}\right)|k^0\ra \otimes
\exp\left(\sum_{n=1}^\infty
\sum_{s=1}^{25} (-1)^{d_s} \, {1\over
n} \, \alpha^s_{-n} \bar
\alpha^s_{-n} \right) |\vec k_\parallel=0,
\vec k_\perp\ra\, \nonumber \\
&& \qquad \qquad \otimes
\exp\left(-\sum_{n=1}^\infty (\bar b_{-n} c_{-n} +
b_{-n}
\bar c_{-n})
\right) (c_0+\bar c_0)c_1\bar c_1 |0\ra\,
\bigg]\nonumber \\
&& = 0\, ,
\een
for any $k^0$ and $\vec k_\perp$.
Since $|\BB\ra=|\BB_1\ra + |\BB_2\ra$ is
BRST invariant, and $|\BB_1\ra$ is BRST invariant, we see that $|\BB_2\ra$ 
is also BRST invariant:
\be \label{eb2gi}
(Q_B+\bar Q_B)|\BB_2\ra=0\, .
\ee
This also follows directly from the fact that $L^{X^0}_n - \bar 
L^{X^0}_{-n}$ annihilates $|\wh\BB\ra_{c=1}$.

Let us define $\wh A_N$ to be an operator 
of level $(N,N)$, composed of negative moded oscillators of $X^0$, $X^s$, 
$b$, $c$, $\bar b$ and $\bar c$ such that
\be \label{eexpan}
\exp\left[\sum_{n=1}^\infty \, \left( -{1\over n}
\alpha^0_{-n}\bar\alpha^0_{-n} + 
\sum_{s=1}^{25} (-1)^{d_s} \, {1\over
n} \,  \alpha^s_{-n} \bar
\alpha^s_{-n} - (\bar b_{-n} c_{-n} +
b_{-n}
\bar c_{-n}) \right)
\right] = \sum_{N=0}^\infty \wh A_N\, .
\ee
Here $\wh A_0=1$.
Then $|\BB_1\ra$ can be expressed as 
\be \label{ebb1exp}
|\BB_1\ra = \TT_p \, \int {d^{25-p} k_\perp \over (2\pi)^{25-p}} \,
\sum_{N=0}^\infty \, \wh A_N
\,
(c_0+\bar c_0) \, c_1 \, \bar c_1 \, \wt f\left(X^0(0)\right) \, 
|k^0=0, \vec k_\parallel=0, \vec k_\perp\ra\, .
\ee
Also since $Q_B$ and $\bar Q_B$ preserves the 
level of a state, \refb{estrong} and \refb{eexpan} give
\be \label{eqcond}
(Q_B+\bar Q_B) \, \wh A_N \, (c_0+\bar c_0) \, c_1\, \bar c_1\,
|k^0, \vec
k_\parallel=0, \vec k_\perp\ra = 0\, .
\ee
We shall make use of these relations later.

{}From \refb{epriform}, \refb{esecondform} we see that $|\wt\BB\ra_{c=1}$ 
is built on states carrying integer $x$ momentum. 
Upon 
continuation to the 
Minkowski space these correspond to states built on $e^{n X(0)}|0\ra$ for 
integer $n$. 
\refb{ebs4} then allows us to express $|\BB_2\ra$ as 
\be \label{ebb2}
|\BB_2\ra = \TT_p\, \sum_{n=-\infty}^\infty \, 
\sum_{N=1}^\infty \, 
\int {d^{25-p} k_\perp \over 
(2\pi)^{25-p}} \,
\wh\OO^{(n)}_N
\,
(c_0+\bar c_0) \, c_1 \, \bar c_1 \, e^{n X^0(0)} \, |k^0=0, \vec
k_\parallel=0,\vec k_\perp\ra\, ,
\ee
where $\wh\OO^{(n)}_N$ is some fixed combination of negative moded 
oscillators of total
level $(N,N)$. 
The structure of $\wh\OO^{(n)}_N$ is different for different $n$
since the primaries $|j,m\ra$ for $m\ne \pm j$ 
involve different oscillator combinations for different 
$(j,m)$.\footnote{This is apparent from the fact that the level of the 
oscillator combination in $\wh P_{j,m}$ in \refb{epriform} is $(j^2-m^2)$ 
which clearly depends on $j$ and $m$ 
for $m\ne \pm j$.}
Note that the sum over $N$ starts at 1, since in the 
conformal
field theory involving the $X^0$ field
$|\BB_2\ra$ involves Virasoro descendants of
primaries $|j,m\ra$ of level $(j^2-m^2,j^2-m^2)\ge (1,1)$.
Since $(Q_B+\bar Q_B)$ preserves the momenta as well as the level of a 
state,
we can 
conclude from \refb{eb2gi}, \refb{ebb2} that 
\be \label{eqcondb2}
(Q_B+\bar Q_B) \, \wh \OO^{(n)}_N \, (c_0+\bar c_0) \, c_1\, \bar c_1\,
e^{nX^0(0)} \, |k^0=0, \vec
k_\parallel=0, \vec k_\perp\ra = 0\, .
\ee

We now note some crucial differences between the structure of $|\BB_1\ra$ 
and 
$|\BB_2\ra$. Since the function $\wt f(x^0)$ defined in \refb{efdef} 
vanishes as $x^0\to \pm\infty$, the source terms for the closed string 
fields produced by $|\BB_1\ra$ vanish asymptotically. In contrast the 
source associated with the $n$-th term in the sum in
\refb{ebb2} is proportional to $e^{n x^0}$, and grows for 
$x^0\to\infty$ ($x^0\to -\infty$) for positive (negative) $n$. The other 
crucial difference between $|\BB_1\ra$ and $\BB_2\ra$ is that while the 
requirement of 
BRST invariance does not give any constraint on the time dependence of 
the source terms generated by $|\BB_1\ra$, it completely fixes the time 
dependence of the source terms generated by $|\BB_2\ra$. To see this we 
note that if we replace $\wt f(x^0)$ by any arbitrary function in the 
expression \refb{ebs3} of $|\BB_1\ra$, we shall still get a BRST invariant 
state due to eq.\refb{estrong}. On the other hand, if we replace the 
factor $e^{nX^0(0)}$ in the 
expression \refb{ebb2} of $|\BB_2\ra$ by an arbitrary function 
$g^{(n)}(X^0(0))$, $|\BB_2\ra$ ceases to be BRST invariant.
To see this we use eqs.\refb{esecondform}, \refb{ehbdef}, 
\refb{ebs4} to express $|\BB_2\ra$ in a form 
slightly different from that given in \refb{ebb2}:
\be \label{ediff1}
|\BB_2\ra = \TT_p\, \sum_{j=1}^\infty \, \sum_{m=-(j-1)}^{j-1} \, \int 
{d^{25-p} k_\perp \over
(2\pi)^{25-p}} \, f_{j,m} (\tl) \, \wh\RR_{j,m} \, 
(c_0+\bar c_0)c_1\bar c_1 e^{2 m X^0(0)} \, |k^0=0, \vec
k_\parallel=0,\vec k_\perp\ra\, ,
\ee
where 
\be \label{ediff2}
\wh\RR_{j,m} = \wh \NN_{j,m} \, \exp\left(\sum_{n=1}^\infty
\sum_{s=1}^{25} (-1)^{d_s} \,{1\over
n} \,  \alpha^s_{-n} \bar
\alpha^s_{-n} \right) \, \exp\left(-\sum_{n=1}^\infty (\bar b_{-n} c_{-n} 
+
b_{-n}
\bar c_{-n})
\right)\, .
\ee
$\wh \NN_{j,m}$ in turn is an operator made of $\alpha^0_{-n}$, 
$\bar\alpha^0_{-n}$ for $n>0$  
such that the Ishibashi state $|j,m\ra\ra$ in the Minkowski theory is 
given by:
\be \label{ediff3}
|j,m\ra\ra = \wh \NN_{j,m} \, e^{2m X^0(0)} \, |0\ra\, .
\ee
{}From eqs.\refb{ediff2}, \refb{ediff3} it is clear that $\wh \RR_{j,m}$ 
does not have 
any explicit $\tl$ dependence. 
Now consider generalizing the source terms given by $|\BB_2\ra$ in a way 
that preserves the operator structure but gives the source terms arbitrary 
time dependence:
\be \label{ediff4}
|\BB_2\ra' = \TT_p\, \sum_{j=1}^\infty \, \sum_{m=-(j-1)}^{j-1} \, \int
{d^{25-p} k_\perp \over
(2\pi)^{25-p}} \, \wh\RR_{j,m} \,
(c_0+\bar c_0)c_1\bar c_1 g_{j,m} (X^0(0)) \, |k^0=0, \vec
k_\parallel=0,\vec k_\perp\ra\, .
\ee
Requiring the BRST invariance of $|\BB_2\ra'$ 
\be \label{ediff5}
(Q_B+\bar Q_B)|\BB_2\ra' = 0\, ,
\ee
we get
\be \label{ediff6}
\p_0 \left(e^{-2m x^0}  g_{j,m}(x^0)\right) = 0\, .
\ee
This follows from the fact that for generic $g_{j,m}(x^0)$, the state 
$\wh\PP_{j,m} \, g_{j,m} (X^0(0)) \, |0\ra$ is no longer a primary state, 
and more generally $\wh \NN_{j,m} \, g_{j,m} (X^0(0)) \, |0\ra$ is no 
longer an 
Ishibashi state. Thus there will be additional contributions from the 
$(c_{-n} L^{X^0}_n+\bar c_{-n} \bar L^{X^0}_{n})$ 
terms in $Q_B+\bar Q_B$ acting on this state. These additional 
contributions will 
vanish when $g_{j,m}(x^0)$ satisfy \refb{ediff6}.

The general arguments given in refs.\cite{0203265,0309074} as well as the 
explicit form of 
eq.\refb{ediff6} suggests that
$e^{-2m x^0}  g_{j,m}(x^0)$ can be thought of as a conserved 
charge.\footnote{Note that for each pair $(j,m)$ with $j\ge 1$, $-(j-1)\le 
m \le (j-1)$, we can in principle define a conserved charge for every 
primary of the $c=25$ CFT, since the action of $(Q_B+\bar Q_B)$ does not 
mix the Verma modules built over such primaries. However all 
these charges will be proportional to $f_{j,m}(\tl)$.} {}From this we see 
that the conserved charges are characterized by two 
half integers $(j,m)$ in the range $j\ge 1$, $-(j-1)\le m \le (j-1)$. 
Comparing \refb{ediff1} and \refb{ediff4} we see that
for the boundary state $|\BB_2\ra$ associated with the rolling 
tachyon solution, we have
\be \label{ediff7}
e^{-2m x^0}  g_{j,m}(x^0) = f_{j,m}(\tl)\, .
\ee
\refb{efjm2} shows that all these charges vanish at $\tl={1\over 2}$. This 
is expected since $\tl={1\over 2}$ describes the closed string vacuum 
without any D-brane. For $j=1$, $m=0$ the conserved charge is proportional 
to the energy 
density of the D-brane\cite{0203265}.

We end this section with a cautionary remark. The analysis given here 
shows that the charges $g_{j,m}(x^0)$ are conserved at least in a 
subsector of the open string theory which corresponds to adding boundary 
perturbation involving $X^0$ and its derivatives, since in this case the 
final boundary state will have the product structure $|\BB\ra_{c=1} 
\otimes |\BB\ra_{c=25} \otimes |\BB\ra_{ghost}$, with $|\BB\ra_{c=1}$ 
given by some linear combination of the Ishibashi states in the $c=1$ 
conformal field theory. Thus $g_{j,m}(x^0)$ can be defined and shown to be 
conserved following the procedure outlined in this section. Whether these 
conservation laws have analogs in the full open string theory remains to 
be 
seen. Nevertheless having conservation laws of this type even in a 
restricted subsector of the theory could facililate analysis of 
classical solutions in that subsector. In section \ref{smatrix} we 
shall see that in the 
two dimensional string theory these conservation laws do hold in the full 
theory.

\sectiono{Closed String Field Produced by the Rolling Tachyon in Critical 
String Theory} \label{sclosed}

The closed string field $|\Psi_c\ra$ is a state of ghost number 2 in the 
Hilbert space of first quantized closed string theory, satisfying the 
constraint\cite{9705241}
\be \label{efg1.5}
b_0^-|\Psi_c\ra = 0\, , \qquad L_0^-|\Psi_c\ra = 0\, ,
\ee
where we define
\be \label{efg2.5}
c_0^\pm = (c_0 \pm \bar c_0), \qquad b_0^\pm = (b_0 \pm \bar b_0)\, ,
\qquad
L_0^\pm=(L_0 \pm \bar L_0)\, .
\ee
$c_n$, $\bar c_n$, $b_n$, $\bar b_n$ are the
usual ghost oscillators, and $L_n$, $\bar L_n$ are the total Virasoro
generators. The quadratic part of the closed string field theory action is
given by:
\be \label{efg2}
-{1\over K g_s^2}\la \Psi_c|c_0^- (Q_B+\bar Q_B) |\Psi_c\ra\, ,
\ee
where $Q_B$ and $\bar Q_B$ are the holomorphic and
anti-holomorphic components of the BRST charge, $K$ is a normalization
constant to be given in eq.\refb{eknorm}, and $g_s$ is the closed
string coupling constant. In the
presence of
the
D-brane we need to add an extra source term to the action:
\be \label{efg3}
\la \Psi_c| c_0^- |\BB\ra\, .
\ee
The equation of motion of
$|\Psi_c\ra$ is then
\be \label{efg4}
2\, (Q_B+\bar Q_B)\, |\Psi_c\ra = K\, g_s^2 \, |\BB\ra\, .
\ee
Clearly by a rescaling of $|\Psi_c\ra$ we can change $K$ and the
normalization of $|\BB\ra$. However once the normalization of $|\BB\ra$ is
fixed in a convenient manner (as in eq.\refb{efg10}), the normalization 
constant $K$ can be
determined by requiring that in the euclidean theory the classical action, 
obtained after
eliminating $|\Psi_c\ra$ using its equation of motion \refb{efg4} and 
substituting it back in the sum of \refb{efg2} and \refb{efg3},
reproduces the one loop partition function
$Z_{open}$ of the open string theory on the D-brane. Using the solution to 
\refb{efg4} given in \refb{es9.1} we get
\be \label{eknorm}
Z_{open} = {1\over 2} \, K\, g_s^2 \, \la \BB| (c_0-\bar c_0) \, [2(L_0 + 
\bar 
L_0)]^{-1} \, (b_0+\bar b_0) \,
|\BB\ra\, .
\ee
Since $Z_{open}$ is independent of $g_s$, and $|\BB\ra$ is inversely 
proportional to $g_s$ due to the $\TT_p$ factor in \refb{efg10}, we see 
that $K$ is a purely numerical constant.

We want to look for solutions to  eq.\refb{efg4}.
Noting that $|\BB\ra$ is BRST invariant,
and that $\{Q_B+\bar Q_B, b_0+\bar b_0\} = (L_0 + \bar L_0)$, we can write
down a
solution to equation \refb{efg4} as:
\be \label{es9.1}
|\Psi_c\ra = K\, g_s^2 \, [2(L_0 + \bar L_0)]^{-1} \, (b_0+\bar b_0) \,
|\BB\ra\, .
\ee
This solution satisfies the Siegel gauge condition $(b_0+\bar
b_0)|\Psi_c\ra=0$. We can of course construct other solutions which are
gauge equivalent to this one by adding to $|\Psi_c\ra$ terms of the form
$(Q_B+\bar Q_B)|\Lambda\ra$. However even within Siegel gauge, the right
hand side of \refb{es9.1} is not defined
unambiguously.
Since free closed string field theory in Minkowski space has infinite
number of
plane
wave solutions in the Siegel gauge, satisfying
\be \label{esiegel}
(L_0+\bar L_0)|\Psi_c\ra
= 0\, , \qquad (b_0+\bar b_0)|\Psi_c\ra
= 0 \, ,
\ee
the right hand side of \refb{es9.1} is defined only up to addition of 
solutions of \refb{esiegel}. However, since a 
solution of \refb{esiegel}
does not in general satisfy the full set of source free string field field 
equations
of motion
$(Q_B+\bar Q_B)|\Psi_c\ra=0$, addition of an arbitrary solution of 
\refb{esiegel} to a solution to \refb{efg4} will not, in general, 
generate a solution of \refb{efg4}. Thus we need to carefully choose a
prescription for defining the right hand side of \refb{es9.1} in order to
construct
a solution of eq.\refb{efg4}.
A natural prescription (known as the
Hartle-Hawking prescription) is to begin with the
solution of the associated equations of motion in the Euclidean theory
where there is a unique solution to eq.\refb{es9.1} (which therefore
satisfies the full equation \refb{efg4}) and then analytically continue
the
result to the Minkowski space along the branch passing through the 
origin $x^0=0$\cite{0303139,0304192,0402124}. This is the prescription we 
shall 
follow.

Since the boundary state $|\BB\ra$ for the rolling tachyon configuration 
can be regarded as a sum of two 
components $|\BB_1\ra$ and $|\BB_2\ra$ each of which are separately gauge 
invariant, we shall analyze their effects separately. Let us denote by 
$|\Psi_c^{(1)}\ra$ and $|\Psi_c^{(2)}\ra$ the closed string field 
configurations produced by $|\BB_1\ra$ and $|\BB_2\ra$ respectively. We 
begin with the analysis of
$|\Psi_c^{(1)}\ra$. The result for this is already contained implicitly
in 
\cite{0303139,0304192,0402124}, but we shall reproduce these results 
for 
completeness. 
Let $\phi_n(\vec k, x^0)$ denote a closed string field with spatial
momenta $\vec k$,
appearing in the expansion of
$|\Psi^{(1)}_c\ra$ as the coefficient of a state for which the oscillator 
contribution to the $L_0+\bar L_0$ eigenvalue is $m_n^2/2$. We shall call 
$m_n$ the mass of $\phi_n$ even though $\phi_n$ may not represent a 
physical closed string state of mass $m_n$. Let $j_n(\vec k, x^0)$ be 
the source of $\phi_n(\vec k, x^0)$ 
appearing in the expansion of $(b_0+\bar b_0)\,|\BB_1\ra$.  
Since
$2(L_0+\bar
L_0)$ acting on $|\Psi_c\ra$ has the effect of converting
$\phi_n(\vec
k, x^0)$ to $(\p_0^2 + \vec k^2 + m_n^2)\phi_n(\vec
k, x^0)$,  the
equations of motion satisfied by this field is given by:
\be \label{es9.3}
(\p_0^2 + \vec k^2 + m_n^2)\phi_n(\vec k, x^0) = K\, g_s^2 \, j_n(\vec k, 
x^0)\, .
\ee
In the euclidean theory, obtained by replacing $x^0$ by $ix$, the equation
takes the form:
\be \label{es9.4}
(-\p_x^2 + \vec k^2 + m_n^2) \, \phi_n(\vec k, ix) = K\, g_s^2 \, j_n(\vec 
k, ix)\, .
\ee
This has solution:
\be \label{es9.8}
\phi_n(\vec k, ix) = {K\, g_s^2\over 2\omk} \left[ \int_{-\infty}^x 
e^{-\omk
(x-x')} j_n(\vec k, ix') dx' + \int_x^\infty e^{\omk (x-x')} j_n(\vec
k, ix') dx'
\right] \, .
\ee
where $\omk = \sqrt{\vec k^2 + m_n^2}$. 
In terms of the variable $x^0=ix$, $x^{\prime 0} = i x'$, this may be
written as
\ben \label{es9.9}
\phi_n(\vec k, x^0) &=& -{i \, K\, g_s^2\over 2\omk} \left[ 
\int_{-i\infty}^{x^0}
e^{ i \omk (x^0 - x^{\prime 0})} j_n(\vec k, x^{\prime 0}) d x^{\prime 0}
+ \int_{x^0}^{i\infty} e^{-i \omk (x^0 - x^{\prime 0})} j_n(\vec
k, x^{\prime
0}) d x^{\prime 0} \right] \nonumber \\
&=& {i \, K\, g_s^2\over 2\omk}  \left[\int_{i\infty}^{x^0} e^{-i \omk 
(x^0 -
x^{\prime 0})} j_n(\vec k, x^{\prime
0}) d x^{\prime 0} - \int_{-i\infty}^{x^0}
e^{ i \omk (x^0 - x^{\prime 0})} j_n(\vec k, x^{\prime 0}) d x^{\prime
0}\right]
\, , \nonumber \\
\een
for $x^0$ lying on the imaginary axis. We now define its analytic
continuation to the real axis by analytically continuing $\phi_n(\vec k,
x^0)$ near the origin along the real axis. This gives, for real $x$:
\be \label{es9.10}
\phi_n(\vec k, x^0) = {i \, K\, g_s^2\over 2\omk}  \left[\int_C \, e^{-i 
\omk 
(x^0 -
x^{\prime 0})} j_n(\vec k, x^{\prime
0}) d x^{\prime 0} - \int_{C'} e^{ i \omk (x^0 - x^{\prime 0})}
j_n(\vec k, x^{\prime 0}) d x^{\prime 0}\right]
\, ,
\ee
where the contour $C$ runs from $i\infty$ to the origin along the
imaginary $x^{\prime 0}$ axis, and then to $x^0$ along the real $x^{\prime
0}$ axis, and the contour $C'$ runs from $-i\infty$ to the origin along
the
imaginary $x^{\prime 0}$ axis, and then to $x^0$ along the real $x^{\prime
0}$ axis. These are known as the Hartle-Hawking contours.

The specific form of $j_n(\vec k, x^{0})$ can be read out from the 
expansion \refb{ebb1exp} of $|\BB_1\ra$. The leven $(N,N)$ term acts as a 
source for a closed string field of mass$^2=4(N-1)\equiv m_N^2$. 
Then using \refb{ebb1exp} and \refb{es9.10} we can 
express the closed string field $|\Psi^{(1)}_c\ra$ produced by $|\BB_1\ra$ 
as:
\be \label{esa0}
|\Psi_c^{(1)}\ra = 2\, K\, g_s^2 \, \TT_p \, \int {d^{25-p} k_\perp\over
(2\pi)^{25-p}} \,
\sum_{N\ge 0} \, \wh A_N \, h^{(N)}_{\vec k_\perp}(X^0(0)) \, c_1\,
\bar c_1\, |k_\parallel=0, \vec k_\perp\ra\, ,
\ee
where
\be \label{esa-1}
h^{(N)}_{\vec k_\perp}(x^0)
= {i \over 2\onk} \, \left[\int_C \, e^{-i \onk
(x^0 -
x^{\prime 0})} \wt f(x^{\prime
0}) d x^{\prime 0} - \int_{C'} e^{ i \onk (x^0 - x^{\prime 0})}
\wt f(x^{\prime 0}) d x^{\prime 0}\right]
\, ,
\ee
with
\be \label{defonk}
\onk=\sqrt{\vec k_\perp^2 + m_N^2} = \sqrt{\vec k_\perp^2 + 4(N-1)}\, .
\ee
The overall multiplicative factor of 2 in \refb{esa0} is due to the factor 
of 2 produced by the anti-commutator of $(b_0+\bar b_0)$ and $(c_0+\bar 
c_0)$ in $(b_0+\bar b_0)|\BB_1\ra$.
In the $x^0\to\infty$ limit we can evaluate the integrals by closing the 
contours $C$ and $C'$ in the first and the
fourth quadrangles
respectively. This gives:
\be \label{es9.17aa}
h^{(N)}_{\vec k_\perp}(x^0\to\infty) =
{\pi\over
\sinh(\pi\onk)} \, {1\over 2\onk} \, \left[ e^{-i\onk (x^0 + \ln
\sin(\pi\tl)) } +  e^{i\onk (x^0 + \ln
\sin(\pi\tl)) } \right] \, .
\ee
Substituting this into \refb{esa0} we get the asymptotic form of 
$|\Psi_c^{(1)}\ra$ in the $x^0\to\infty$ limit to be:
\ben \label{es9.17bb}
|\Psi_c^{(1)}\ra &\to& 2\, K\, g_s^2 \, \TT_p \, \int {d^{25-p} 
k_\perp\over
(2\pi)^{25-p}} \, \sum_{N\ge 0} \, {\pi\over
\sinh(\pi\onk)} \, {1\over 2\onk} \, \wh A_N\,  c_1\, \bar c_1\, \,
\nonumber \\
&& \left[ e^{-i\onk \, \ln
\sin(\pi\tl)} |k^0=\onk, \vec k_\parallel=0, \vec k_\perp\ra +  e^{i\onk
\, \ln
\sin(\pi\tl)} |k^0=-\onk, \vec k_\parallel=0, \vec k_\perp\ra\right] \, .
\nonumber \\
\een

$|\Psi_c^{(1)}\ra$ defined in \refb{esa0} clearly satisfies the Siegel 
gauge equations of motion:
\be \label{erestrict}
2(L_0+\bar L_0) |\Psi_c^{(1)}\ra = K\, g_s^2 \, (b_0+\bar b_0) |\BB_1\ra\, 
.
\ee
Let us now try to verify explicitly that \refb{esa0} satisfies the full 
set of 
equations of motion \refb{efg4} with $|\BB\ra$ replaced by $|\BB_1\ra$. 
For this we express $(Q_B+\bar Q_B)$ as a sum of two terms:
\be \label{eqbbreak}
Q_B+\bar Q_B = (c_0 L_0 + \bar c_0 \bar L_0) + \wh Q\, ,
\ee
where $\wh Q$ does not contain any $c$ or $\bar c$ zero modes and hence 
anti-commute with $b_0$ and $\bar b_0$. \refb{eqcond} then gives: 
\be \label{eqc1}
\wh Q \,  \wh A_N \, (c_0+\bar c_0) \, c_1\, \bar c_1\,
|k^0, \vec
k_\parallel=0, \vec k_\perp\ra = 0\, ,
\ee
where we have used the fact that $(L_0-\bar L_0)$ annihilates $ \wh A_N \, 
(c_0+\bar c_0) \, c_1\, \bar c_1\,
|k^0, \vec
k_\parallel=0, \vec k_\perp\ra$. Applying $(b_0+\bar b_0)$ on \refb{eqc1} 
and using the fact that $(b_0+\bar b_0)$ anti-commutes with $\wh Q$ and 
commutes with $\wh A_N$, we 
get:
\be \label{eqc5}
\wh Q  \wh A_N \, c_1\, \bar c_1\,
|k^0, \vec
k_\parallel=0, \vec k_\perp\ra = 0\, .
\ee
Using \refb{erestrict} - \refb{eqc5} we get:
\be \label{efull}
2\, (Q_B+\bar Q_B) |\Psi_c^{(1)}\ra = 2\, (c_0 L_0 + \bar c_0 \bar L_0) 
|\Psi^{1)}\ra = K\, g_s^2 \, |\BB_1\ra\, ,
\ee
as required.

The crucial relation leading to the final result is \refb{eqc5} which 
shows that $\wh Q$ annihilates the closed string field 
configuration $|\Psi^{(1)}_c\ra$. This in turn is a consequence of the 
fact that 
for a given 
momentum $(k^0, \vec k)$ and a given level $(N,N)$ the 
combination of oscillators $\wh A_N$ that appears in the expression for 
$|\Psi^{(1)}\ra$ is the same as the one that appears in the expression for 
$|\BB_1\ra$. This is a special property of the specific definition of 
$(L_0+\bar L_0)^{-1}$ through the Hartle-Hawking prescription that we have 
used and will not hold for a generic 
definition.\footnote{Of course this does not mean that this is the only 
possible prescription.} For example, we could have expressed the 
level $N$ 
contribution to $|\BB_1\ra$ as a linear combination of some fixed basis of 
level $(N,N)$ states, and used different prescription for $(L_0+\bar 
L_0)^{-1}$ for these different basis states ({\it e.g.} Hartle-Hawking 
prescription for some and retarded Greens function for the others). The 
result will be a level $(N,N)$ state that involves an oscillator 
combination different from $\wh A_N$, and would not be 
annihilated by $\wh Q$.

Since $\wt f(x^0)$ vanishes as $x^0\to\pm\infty$, in the far future and 
far past we
are left with pure closed string background satisfying free field
equations of motion $(Q_B+\bar Q_B)|\Psi_c^{(1)}\ra=0$, {\it i.e.}
on-shell
closed string field
configuration. One amusing point to note is that 
\refb{es9.17bb} does not vanish even in
the $\tl\to {1\over 2}$ limit, even though the boundary state $|\BB_1\ra$
vanishes in this limit\cite{0304192}. This is because in the euclidean 
theory the
boundary state $|\BB_1\ra$ represents an array of D-branes with Dirichlet
boundary condition on $X=-i X^0$, located at $x=(2n+1)\pi$. This produces
a non-trivial background in the euclidean theory, which, upon inverse Wick
rotation, produces a source free closed string background in the Minkowski
theory\cite{0304192,0402124}. The other important point to note is that 
the 
dependence of $|\Psi^{(1)}_c\ra$ on $\tl$ in the $x^0\to\infty$ limit 
comes only through a $\tl$ dependent time delay of 
$-\ln(\sin(\pi\tl))$\cite{0303139}.

We now turn to the analysis of closed string fields generated by
$|\BB_2\ra$. We begin with the form \refb{ebb2} of $|\BB_2\ra$. 
Since $ \wh\OO^{(n)}_N
\,
(c_0+\bar c_0) \, c_1 \, \bar c_1 \, e^{n X^0(0)} \, |\vec k_\perp\ra$ is
an eigenstate of $2(L_0+\bar L_0)$ with eigenvalue $(4(N-1)+n^2 + \vec
k_\perp^2)$, the natural choice of the closed string field produced by
$|\BB_2\ra$, as given in eq.\refb{es9.1}, is
\ben \label{ebb2cl}
|\Psi_c^{(2)}\ra &=& 2 \,  K\, g_s^2 \, \TT_p \, \sum_{n\in Z} \, 
\sum_{N=1}^\infty 
\,  \int \,{ d^{25-p}
k_\perp \over (2\pi)^{25-p}} \,
\left(4(N-1)+n^2 + \vec
k_\perp^2\right)^{-1} \nonumber \\
&& \qquad \qquad \qquad \qquad \qquad \, \wh \OO^{(n)}_N
\, c_1 \, \bar c_1 \, e^{n X^0(0)} \, |k^0=0, \vec k_\parallel=0,\vec
k_\perp\ra\, .
\een
Clearly, this is the result that we shall get if we begin with the 
closed
string background produced by the boundary state in the euclidean theory
and then analytically continue it to the Minkowski space. 
Following the same procedure as in the case of $|\Psi^{(1)}_c\ra$ one can 
show that \refb{ebb2cl} satisfies the full string field equation:
\be \label{efull2}
2\, (Q_B+\bar Q_B) |\Psi_c^{(2)}\ra = 2\, (c_0 L_0 + \bar c_0 \bar L_0) 
|\Psi^{(2)}\ra = K\, g_s^2 \, |\BB_2\ra\, .
\ee
The crucial relation that establishes the first equality in \refb{efull2} 
is:
\be \label{ecruc}
\wh Q \, \wh \OO^{(n)}_N
\, c_1 \, \bar c_1 \, |k^0, \vec k_\parallel=0,\vec
k_\perp\ra\, ,
\ee
which follows from \refb{eqcondb2}.

The space-time
interpretation of this state for a given value of $n$ is that it
represents a field which grows as $e^{n x^0}$. For positive $n$ this
diverges as $x^0\to\infty$ and for negative $n$ this diverges as
$x^0\to-\infty$.
On the other hand in the transverse spatial directions the solution falls
off as
$G(\vec x_\perp, \sqrt{4(N-1) +n^2})$ where $G(\vec x_\perp, m)$ denotes
the Euclidean Greens function of a scalar field of mass $m$ in $(25-p)$
dimensions. Since $G(\vec x_\perp, m) \sim e^{-m |\vec x_\perp|}/ |\vec
x_\perp|^{(24-p)/2}$ for non-zero $m$ and large $|\vec x_\perp|$, we see
that the coefficients of the closed string field associated with the
state $\wh \OO^{(n)}_N c_1\bar c_1|k\ra$ behaves as
\be \label{eexpon}
\exp\left(n x^0 - \sqrt{4(N-1) +n^2} \, |\vec x_\perp|\right)/ |\vec
x_\perp|^{(24-p)/2}\, .
\ee
Thus at any given time $x^0$, the field associated with $\wh \OO^{(n)}_N
c_1\bar c_1|k\ra$ is small for $|\vec x_\perp|>> n x^0/\sqrt{4(N-1) +n^2}$
and large for $|\vec x_\perp| <<  n x^0/\sqrt{4(N-1) +n^2}$. We can
view such a field configuration as a disturbance
propagating outward in
the transverse directions from
$\vec x_\perp=0$ at a speed of $ n/\sqrt{4(N-1) +n^2}$. Since $N\ge 1$,
this
is less than the speed of light but approaches the speed of light for
fields for which $N<< n^2$.

Since the source for the closed string fields produced by $|\BB_2\ra$
is localized at $\vec x_\perp=0$, $|\Psi^{(2)}_c\ra$ should satisfy source
free closed
string
field equations away from the origin. It is easy to see that 
in the position space representation \refb{ebb2cl}
is annihilated by $(L_0+\bar L_0)$ away from $\vec x_\perp=0$. 
Eq.\refb{ecruc} then establishes that 
$|\Psi^{(2)}_c\ra$ satisfy the full set of free field equations of
motion:
$(Q_B+\bar
Q_B)|\Psi^{(2)}_c\ra=0$ away from $\vec x_\perp=0$.

Using \refb{efjm2} we see that $|\wh\BB\ra_{c=1}$ and hence $|\BB_2\ra$ 
vanishes for $\tl={1\over 2}$. As a result the operators $\OO^{(n)}_N$ 
defined
through \refb{ebb2} vanish, and hence $|\Psi_c^{(2)}\ra$ given in
\refb{ebb2cl} also vanishes. Thus in the $\tl\to {1\over 2}$ limit the
$|\Psi_c^{(1)}\ra$ given in \refb{es9.17bb} is the only contribution to
the closed string background. This of course is manifestly finite in the
$x^0\to\infty$ limit.

For D0-branes, and more generally for D-$p$-branes with all tangential 
directions compactified on a torus, $|\Psi^{(1)}_c\ra$ represents a 
collection of massive, non-relativistic closed 
strings\cite{0303139,0304192}. The total energy 
density stored in $|\Psi^{(1)}_c\ra$ turns out to be infinite. Naively 
this would suggest that the backreaction due to closed string emission 
effects invalidate the classical open string analysis of the system. 
However an alternative interpretation suggested in 
\cite{0305011,0306132,0306137,0308068,0402027} 
is 
that $|\Psi_c^{(1)}\ra$ gives the dual closed string representation of the 
tachyon matter 
predicted by the tree level open string analysis\cite{0203211,0203265}. 
The evidence for this comes from the fact that the 
closed string field configuration described by $|\Psi_c^{(1)}\ra$ turns 
out 
to have properties similar to the tachyon matter predicted by the 
classical open string analysis.  We believe that a 
similar interpretation must exist also for the $|\Psi^{(2)}_c\ra$ 
component of the closed string background for a generic $\tl$, but the 
lack of a complete 
understanding of the tree level open string results prevents us from 
arriving at this understanding at present. In the next few sections we 
shall 
address the same problem in two dimensional string theory where a complete 
understanding of the tree level open string results are available through 
the matrix model description of the system. 

Before concluding this section we must mention that there is a different 
approach to constructing the boundary state of the rolling tachyon 
solution that gets rid of the exponentially
growing
terms produced by $|\BB_2\ra$\cite{0301038}. In this approach we
begin with an appropriate boundary state in the Liouville theory with 
central charge $>1$ where the discrete higher level primares are 
absent\cite{0306026}, analytically continue the result to the Minkowski 
space, and then take the $c\to 1$ limit.
The final boundary state we arrive at this way has the form:
\be \label{emin1}
|\BB\ra_{c=1} = \int d E F(E) |E\ra\ra + \cdots\, ,
\ee
where $F(E)\propto e^{-iE\ln(\sin(\pi\tl))} / sinh(\pi E)$ is the Fourier 
transform of $\wt
f(x^0)$, $|E\ra\ra$ is the Ishibashi state built on the primary 
$|E\ra=\exp(-iE X^0(0))|0\ra$, 
and $\cdots$ denotes the contribution from the Ishibashi states on higher
level primaries at $E=0$. Clearly this form of the boundary state does not
have any exponentially growing contribution. While this could be the 
correct prescription, in this paper we have chosen 
to
proceed with the prescription of computing the boundary state as well as 
the closed string fields produced by the boundary state
in the Euclidean $c=1$ theory, 
and then analytically continuing it to the Minkowski theory along the 
branch 
passing through the origin of the (complex) time coordinate $x^0$.

\sectiono{Boundary State for the Rolling Tachyon in Two Dimensional String 
Theory} \label{s2d}

So far our discussion has taken place in the context of critical bosonic 
string theory. In this section we
shall study the boundary state for the rolling tachyon system in two 
dimensional string theory.
We begin by reviewing the bulk conformal field theory associated with the
two dimensional string theory. The world-sheet action of this
CFT is given by the sum of three separate components:
\be \label{e11.1}
s = s_L + s_{X^0} + s_{ghost}\, ,
\ee
where $s_L$ denotes the Liouville field theory with central charge 25,
$s_{X^0}$ denotes the conformal field theory of a single scalar field
$X^0$ describing the time coordinate
and 
$s_{ghost}$ denotes the usual
ghost action involving the fields $b$, $c$, $\bar b$ and $\bar c$. Of
these $s_{X^0}$ and $s_{ghost}$ are familiar objects. The Liouville action
$s_L$ on a flat world-sheet is given by:
\be \label{e11.2}
s_L = \int d^2 z \left(\p_z\vp \p_{\bar z} \vp + \mu e^{2\vp}\right)
\ee
where $\vp$ is a world-sheet scalar field and $\mu$ is a constant
parametrizing the theory. 
We shall set $\mu=1$ by shifting $\vp$ by ${1\over 2} \ln\mu$.
The
scalar field $\vp$ carries a background charge $Q=2$ (which is not visible
in
the flat world-sheet action \refb{e11.2} but controls the coupling of
$\vp$ to the scalar curvature on a curved world-sheet), so that the theory
has a central charge
\be \label{e11.3}
c = 1 + 6 Q^2 = 25\, .
\ee
For our analysis we shall not use the explicit world-sheet action 
\refb{e11.2}, 
but
only use the abstract properties of the Liouville field theory described
in \cite{9206053,9403141,9506136,0101152,0104158}. In particular 
the 
property of 
the bulk 
conformal 
field
theory that we shall be using is that it has a one parameter ($P$) family
of primary
vertex operators, labelled as $V_{Q + i P}$, of conformal weight:
\be \label{e11.4}
\left({1\over 4} (Q^2 + P^2), {1\over 4} (Q^2 + P^2)\right) =
\left(1+{1\over 4} P^2, 1+{1\over 4} P^2\right)\, .
\ee
Thus a generic normalizable state of the bulk Liouville field theory is
given by a
linear combination of the secondary states built over the primary
$V_{Q+iP}(0)|0\ra$.

The closed string field $|\Psi_c\ra$ in this two dimensional string
theory is a ghost
number 2 state satisfying \refb{efg1.5} in the combined state space 
of the
ghost, Liouville and $X^0$ field theory. If we expand $|\Psi_c\ra$ as
\be \label{eexppsi}
|\Psi_c\ra = \int {d P\over 2\pi} \, \int {d E\over 2\pi} \, \phi(P, E) \,
c_1 \bar c_1 e^{-i E
X^0(0)} V_{Q+iP}(0) |0\ra + \cdots \, ,
\ee
then $\phi(P,E)$ is the Fourier transform of a scalar field $\phi$ known
as the closed string `tachyon' field.\footnote{Throughout this and the 
next two sections we shall use the same symbol {\it e.g.} $\phi$, to 
denote a field and its Fourier transform with respect to $x^0$ and/or 
$\vp$ coordinates.} Despite its name, it actually
describes a massless particle in this $(1+1)$ dimensional string theory,
since the condition that the state $c_1 \bar c_1 e^{-i E
X^0(0)} V_{Q+iP}(0) |0\ra$ is on-shell is $E^2 - P^2=0$. Thus
physically, $c\, \bar c\, V_{Q + i P}e^{i E X^0}$ may be regarded as the
vertex
operator of a scalar field $\phi$ of momentum $P$ (along the Liouville
direction $\vp$) and energy $E$ in this two dimensional string
theory.
This is the only physical closed
string field in this theory.

We shall normalize $|\Psi_c\ra$ so that its kinetic term is given by:
\be \label{ekinlio}
-\la \Psi_c|c_0^- (Q_B+\bar Q_B) |\Psi_c\ra\, .
\ee
Substituting \refb{eexppsi} into \refb{ekinlio} we see that the kinetic
term for $\phi$ is given by:
\be \label{ekinphi}
-{1\over 2} \, \int {dP\over 2\pi} \, {dE\over 2\pi} \, \phi(-P, -E) (P^2
-
E^2) \phi(P,E)\, .
\ee
Thus $\phi$ has the standard normalization of a scalar field.

The Liouville field theory also has an unstable D0-brane obtained by
putting an appropriate boundary condition on the field $\vp$, and the
usual Neumann boundary condition on the $X^0$ and the ghost fields. Since
$\vp$
is an interacting field, it is more appropriate to describe the
corresponding
boundary CFT associated with the Liouville field
by specifying its abstract properties. The relevant properties are as
follows:
\begin{enumerate}

\item The open string
spectrum in this boundary CFT is described by a single Virasoro module
built
over the SL(2,R) invariant vacuum state.

\item The one point function on the disk of the closed string vertex 
operator $V_{Q +
i P}$ corresponding to this boundary CFT is given
by\cite{0101152,0305159}:
\be \label{e11.5}
\la V_{Q +
i P} \ra_D = {2\ C\over \sqrt \pi} \, i\, \sinh(\pi P) \,
{\Gamma(i P)\over \Gamma(-iP)}\, ,
\ee
where $C$ is a normalization constant to be given in eq.\refb{e11.5cc}.
\end{enumerate}
Since $V_{Q +
i P}$ for any real $P$ gives the complete set of primary states in the
theory, we get the boundary state associated with the D0-brane to 
be:\footnote{The normalization factor of $1/2$ is a reflection of the fact 
that if we take $|\Psi_c\ra=c_1\bar c_1 \, V_{Q+iP}(0)|0\ra$ and calculate 
$\la\Psi_c|c_0^-|\BB\ra$ we get a factor of 2 in the ghost correlator $\la 
0|c_{-1}\bar c_{-1} c_0^- c_0^+ c_1\bar c_1|0\ra$.}
\ben \label{eboulio}
|\BB\ra &=& {1\over2} \, \exp\left(\sum_{n=1}^\infty
{1\over
n} \alpha^0_{-n} \bar
\alpha^0_{-n} \right)|0\ra \, \otimes \, \int {d P\over 2\pi} \, \la V_{Q
-
i P} \ra_D |P\ra\ra \nonumber \\
&& \otimes \, \exp\left(-\sum_{n=1}^\infty
(\bar b_{-n}
c_{-n} +
b_{-n}
\bar c_{-n})
\right) (c_0+\bar c_0)c_1\bar c_1 |0\ra\, ,
\een
where $|P\ra\ra$ denotes the
Ishibashi state in the Liouville theory, built on the primary
$V_{Q+iP}(0)|0\ra$. The normalization constant $C$ is determined
by requiring that if we eliminate $|\Psi_c\ra$ from the combined action
\be \label{ecombined}
-\la \Psi_c|c_0^- (Q_B+\bar Q_B) |\Psi_c\ra + \la \Psi_c|c_0^- |\BB\ra\, ,
\ee
using its equation of motion, then the resulting value of the action
reproduces the one loop open string partition function $Z_{open}$ on the
D0-brane. This gives\cite{0305159}
\be \label{e11.5cc}
C=1\, .
\ee

Given this particular boundary CFT associated with the Liouville field, we
can now combine this with the rolling tachyon boundary CFT associated with
the $X^0$ field to construct a rolling tachyon solution on the D0-brane in
two dimensional string theory.
As in the critical string theory, we divide the boundary state into two
parts, $|\BB_1\ra$ and $|\BB_2\ra$, with
\ben \label{ebb1l}
|\BB_1\ra &=& {1\over 2} \, \exp\left(-\sum_{n=1}^\infty
{1\over
n} \alpha^0_{-n} \bar
\alpha^0_{-n} \right) \wt f(X^0(0)) |0\ra \, \otimes \, \int{dP\over 2\pi}
\,
\la
V_{Q-iP}\ra_D \, |P\ra\ra \nonumber \\
&& \otimes
\exp\left(-\sum_{n=1}^\infty (\bar
b_{-n} c_{-n} +
b_{-n}
\bar c_{-n})
\right) (c_0+\bar c_0)c_1\bar c_1 |0\ra\nonumber \\
&\equiv& {1\over 2} \, \int{dP\over 2\pi} \, \la
V_{Q-iP}\ra_D \, \sum_N \, \wh A_N \, \wt f(X^0(0))
V_{Q+iP}(0)
(c_0+\bar c_0) c_1 \bar c_1 |0\ra\, ,
\een
and
\ben \label{ebb2l}
|\BB_2\ra &=& {1\over 2} \, |\wt\BB\ra_{c=1} \, \otimes \, \int{dP\over 
2\pi} \,
\la
V_{Q-iP}\ra_D \, |P\ra\ra \, \otimes \exp\left(-\sum_{n=1}^\infty (\bar
b_{-n} c_{-n} +
b_{-n}
\bar c_{-n})
\right) (c_0+\bar c_0)c_1\bar c_1 |0\ra\, \nonumber \\
&\equiv& {1\over 2} \, \sum_{n\in Z} \, \sum_{N=1}^\infty \, \int {d P 
\over 2\pi} \, \la V_{Q-iP}  \ra_D
\, \wh\OO^{(n)}_N
\,
(c_0+\bar c_0) \, c_1 \, \bar c_1 \, e^{n X^0(0)} \, V_{Q+iP}(0) \,
|0\ra\, .
\een  
Here $|\wh\BB\ra_{c=1}$ is the inverse Wick rotated version of $|\wt
B\ra_{c=1}$ as defined in eq.\refb{ebs1}, and $\wh A_N$ and 
$\wh\OO^{(n)}_N$ are operators of
level $(N,N)$,
consisting of non-zero mode ghost and $X^0$ oscillators, and the Virasoro 
generators
of the Liouville theory.

As in the case of critical string theory, it is easy to show that 
$|\BB_1\ra$ and $\BB_2\ra$ are separately BRST invariant.

\refb{ebb1l} and \refb{ebb2l} shows that the sources for the various 
closed string fields in the momentum space are proportional to $ \la 
V_{Q-iP}  \ra_D$. It is instructive to see what they correspond to in the 
position space labelled by the Liouville coordinate $\vp$. We concentrate 
on the negative $\vp$ region since
for large negative $\vp$ the effect of the $e^{2\vp}$ term in
\refb{e11.2} is small and the Liouville coordinate behaves like a free
scalar field on the world-sheet. 
Thus $V_{Q+iP}$ takes the form $e^{(Q+i P)\vp}=e^{2\vp + i P \vp}$, and
\be \label{eprr}
|P\ra\ra \sim \wh \OO_L \, e^{2\vp(0) + 
i P\vp(0)} |0\ra\, ,
\ee
where $\wh\OO_L$ is an appropriate operator in the Liouville 
field theory.
In this region the source term becomes 
proportioanl to
\be \label{esource}
\int {d
P\over 2\pi} \, e^{2 \vp + i
P \,
\vp} \la V_{Q-iP}\ra_D \propto
\int {d
P\over 2\pi} \, e^{2 \vp + i
P \,
\vp}  \, \, \sinh(\pi P) \, 
{\Gamma(-i P)\over \Gamma(iP)}\, .
\ee
As it stands the integral is not well defined since $\sinh(\pi P)$ blows
up for large $|P|$. For negative $\vp$, we shall define this integral by
closing the contour in the lower half plane, and picking up the
contribution from all the poles enclosed by the contour. Since the poles 
of
$\Gamma(-iP)$ at $P=-in$ are cancelled by the zeroes of $\sinh(\pi P)$ we 
see that the integrand has no pole in the lower half plane and hence the 
integral
vanishes. Thus the boundary state $|\BB_1\ra$ and 
$|\BB_2\ra$ given in \refb{ebb1l} and \refb{ebb2l} do not produce any 
source term for negative $\vp$. This in turn leads to the identification 
of this 
system as a D0-brane that is localized in the Liouville 
direction\cite{0305159}.

The same argument as in the case of critical string theory indicates that 
$|\BB_2\ra$ encodes information about conserved charges. To see explicitly 
what these conserved charges correspond to, we first express $|\BB_2\ra$ 
in a manner similar to that in \refb{ediff1}
\be \label{ecrit1}
|\BB_2\ra = {1\over 2} \, \sum_{j=1}^\infty \, \sum_{m=-(j-1)}^{j-1} \, 
\int {d P\over 
2\pi} \, \la V_{Q-iP}\ra_D\, f_{j,m} (\tl) \, \wh\RR^{(2d)}_{j,m} \,
(c_0+\bar c_0)c_1\bar c_1 e^{2 m X^0(0)} \, |k^0=0, P\ra\, ,
\ee
where
\be \label{ecrit1.5}
\wh\RR^{(2d)}_{j,m} = \wh \NN_{j,m} \, 
\wh\OO_L\, \exp\left(-\sum_{n=1}^\infty (\bar b_{-n} c_{-n}
+
b_{-n}
\bar c_{-n})
\right)\, ,
\ee
and $f_{j,m}(\tl)$ and $\wh \NN_{j,m}$ are as defined in 
eq.\refb{efjm} and \refb{ediff3} respectively. If we now generalize the 
source term so that it has the 
same operator structure but
arbitrary time dependence:
\be \label{ecrit2}
|\BB_2\ra' ={1\over 2} \, \sum_{j=1}^\infty \, \sum_{m=-(j-1)}^{j-1} \, 
\int {d P\over
2\pi} \, \la V_{Q-iP}\ra_D \, \wh\RR^{(2d)}_{j,m} \,
(c_0+\bar c_0)c_1\bar c_1 \, g_{j,m}(X^0(0)) \, |k^0=0, P\ra\, ,
\ee
then requiring $(Q_B+\bar Q_B)|\BB_2\ra'=0$ gives:
\be \label{ecrit3}
\p_0 \left(e^{-2m x^0}  g_{j,m}(x^0)\right) = 0\, ,
\ee
as in the case of critical string theory. Thus $e^{-2m x^0}  g_{j,m}(x^0)$ 
can be thought of as a conserved charge which takes value $f_{j,m} (\tl)$ 
for $|\BB_2\ra$ given in \refb{ecrit1}.

These charges clearly bear a close relation to the global symmetries of 
this two dimensional string theory discussed in \cite{9108004,9201056} but 
we 
shall not explore this relation here. In section \ref{smatrix} we shall 
directly relate these charges to the conserved charges in the matrix model 
description of this theory, which, in turn, are known to be related to the 
global symmetries discussed in \cite{9108004,9201056}.

\sectiono{Closed String Background Produced by the Rolling Tachyon in Two 
Dimensional String
Theory} \label{s2closed}

We now calculate the closed string field produced by this time dependent
boundary state.
The contribution
from
the $|\BB_1\ra$ part of the boundary state can be easily computed as in
the case of critical string theory, and in the $x^0\to\infty$ limit
takes the form:
\be \label{eb1cont}
|\Psi_c^{(1)}\ra = \int{dP\over 2\pi} \la
V_{Q-iP}\ra_D\, \sum_N \wh A_N \, h^{(N+1)}_{\vec
k_\perp}(X^0(0)) V_{Q+iP}(0)
c_1 \bar c_1 |0\ra\, ,
\ee
where $h^{(N)}_{\vec
k_\perp}(x^0)$ has been defined in \refb{esa-1}. The $(N+1)$ in the
superscript of $h$ in \refb{eb1cont} can be traced to the fact that in
this theory a level $(N,N)$ state has mass$^2=4N$ whereas in the critical
string theory a level $(N,N)$ state had mass$^2=4(N-1)$.

Arguments similar to those given for critical string theory shows that in 
the $x^0\to\infty$ limit
this state is on-shell, {\it i.e.} it is annihilated by the BRST charge
$(Q_B+\bar Q_B)$. Since the only physical states in the theory come from
the
closed string tachyon state, it must be possible to remove all the other
components of $|\Psi_c^{(1)}\ra$ by an on shell gauge transformation of
the
form
$\delta |\Psi_c\ra = (Q_B+\bar Q_B)|\Lambda\ra$ by suitably choosing
$|\Lambda\ra$. Furthermore, since the action of $Q_B$ and $\bar Q_B$ does 
not
mix states of different levels, it must be possible to remove all the
$N>0$ components of $|\Psi_c^{(1)}\ra$ without modifying the $N=0$
component.
Using the expression for $h^{(1)}_{\vec k_\perp}$ from \refb{defonk},
\refb{es9.17aa},
and $\la V_{Q-iP}\ra_D$ from \refb{e11.5} we get
the following expression for the closed string tachyon field
$\phi$ in the $x^0\to\infty$ limit\cite{0305159}
\ben \label{e11.6}
\phi(P, x^0\to\infty) &=& - {\pi\over
\sinh(\pi\omega_P)} \, {1\over 2\omega_P} \, {2\over \sqrt \pi} \, i\,
\sinh(\pi
P) \,
{\Gamma(-i P)\over \Gamma(iP)} \nonumber \\
&& \qquad \, \left[ e^{-i\omega_P (x^0 + \ln
\sin(\pi\tl)) } +  e^{i\omega_P (x^0 + \ln
\sin(\pi\tl)) } \right] \, ,
\een
where $\omega_P = |P|$ since we are dealing with a single massless scalar
particle. This can be simplified as
\be \label{e11.7}
\phi(P, x^0\to \infty) = - i\, {\sqrt\pi\over
P} \,
{\Gamma(-i P)\over \Gamma(iP)}
\, \left[ e^{-i P (x^0 + \ln
\sin(\pi\tl)) } +  e^{i P (x^0 + \ln
\sin(\pi\tl)) } \right] \, .
\ee
This finishes our discussion of closed string radiation induced by the
$|\BB_1\ra$ component of the boundary state. 

We shall now discuss the
closed string background produced by $|\BB_2\ra$.
This can be analyzed
in the same way as in the case of critical string theory. We begin 
with the expression \refb{ebb2l} of
$|\BB_2\ra$.
Since $ \wh\OO^{(n)}_N
\,
(c_0+\bar c_0) \, c_1 \, \bar c_1 \, e^{n X^0(0)} \, V_{Q+iP}(0) \, |0\ra$
in this expression is
an eigenstate of $2(L_0+\bar L_0)$ with eigenvalue $(4N+n^2 +P^2)$, we can
choose the closed string field produced by $|\BB_2\ra$ to be:
\be \label{es12.4b}
|\Psi_c^{(2)}\ra = \sum_{n\in Z} \, \sum_{N=2}^\infty \, \int \,{ dP
\over 2\pi} \, \la V_{Q-iP}  \ra_D \,
\left(4N+n^2 + P^2\right)^{-1}
\, \wh\OO^{(n)}_N
\, c_1 \, \bar c_1 \, e^{n X^0(0)} \, V_{Q+iP}(0) \, |0\ra\, .
\ee
This corresponds to closed string field configurations which grow as
$e^{nx^0}$ for large $x^0$.

A special class of operators among the
$\wh O^{(n)}_N$'s are those which involve only excitations involving the
$\alpha^0$,
$\bar\alpha^0$ oscillators and correspond to higher level {\it primaries}
of the
$c=1$
conformal field theory. As described before, these primaries are 
characterized by SU(2) quantum numbers 
$(j,m)$ with $j\ge 1$, $-j< m < j$, and has dimension $(j^2,
j^2)$. The quantum number $m$ can be identified as $n/2$ in
\refb{es12.4b}. From \refb{epriform}, \refb{ediff3}, \refb{ecrit1}, and 
\refb{ecrit1.5} we see 
that the 
contribution to $|\BB_2\ra$ from these primary states has the form:
\be \label{eb2new}
{1\over 2} \, \sum_{j\ge 1} \, \sum_{m=-j+1}^{j-1} \, \int \,{ dP
\over 2\pi} \, \la V_{Q-iP}  \ra_D \,
f_{j,m}(\tl) \, \wh 
\PP_{j,m} \, (c_0+\bar c_0) \, c_1 
\, \bar c_1 \, e^{2m X^0(0)} \, V_{Q+iP}(0) \, |0\ra\, .
\ee
The level of the operators $\wh\PP_{j,m}$ is
\be \label{espace1}
N = (j^2 - m^2) \, .
\ee
Thus the $|\Psi^{(2)}_c\ra$ produced by this part of $|\BB_2\ra$ takes the 
form:
\be \label{espace2}
|\wc\Psi_c^{(2)}\ra = \sum_{j,m} \, f_{j,m}(\tl) \, \int \,{ dP
\over 2\pi} \, \la V_{Q-iP}  \ra_D
\, (4 j^2 + P^2)^{-1}
\, \wh\PP_{j,m}
\, c_1 \, \bar c_1 \, e^{2m X^0(0)} \, V_{Q+iP}(0) \, |0\ra\, .
\ee

As in the case of critical string theory, it
is instructive to study the behaviour of
$|\Psi^{(2)}_c\ra$ in the position
space characterized by
the Liouville coordinate $\vp$ instead of the momentum space expression
given in \refb{es12.4b}. We concentrate on the negative $\vp$ region since
for large negative $\vp$ the effect of the $e^{2\vp}$ term in
\refb{e11.2} is small and the Liouville coordinate behaves like a free
scalar field on the world-sheet. Let us first focus on the
$|\wc\Psi^{(2)}_c\ra$ part of $|\Psi^{(2)}_c\ra$ as given in 
\refb{espace2}.
In the position space the string field component associated with the state 
$\wh\PP_{j,m}
\, c_1 \, \bar c_1 |E,P\ra$ is given by
\be \label{espace4}
\psi_{j,m}(\vp,x^0) = f_{j,m}(\tl) \, e^{2 m x^0} \, \int {d P\over 
2\pi} 
\, e^{2 \vp + i
P \,
\vp} \, (4 j^2 + P^2)^{-1} \, \la V_{Q-iP}  \ra_D\, .
\ee
Using the expression for $\la V_{Q+iP}  \ra_D$ given in \refb{e11.5} we
get
\be \label{espace5}
\psi_{j,m}(\vp,x^0) = -{2\over \sqrt \pi} \, i\, f_{j,m}(\tl) \, e^{2 
m
x^0} \, \int {d
P\over 2\pi} \, e^{2 \vp + i
P \,
\vp} \, (4 j^2 + P^2)^{-1} \, \sinh(\pi P) \, 
{\Gamma(-i P)\over \Gamma(iP)}\, .
\ee
This integral is not well defined since $\sinh(\pi P)$ blows
up for large $|P|$. As in the analysis of \refb{esource}, for negative 
$\vp$ we shall define this integral by
closing the contour in the lower half plane, and picking up the
contribution from all the poles. Since the poles of
$\Gamma(-iP)$ at $P=-in$ are cancelled by the zeroes of $\sinh(\pi P)$,
the only pole that the integral has in the lower half plane is at
$P=-2ij$. Evaluating the residue at this pole, we get
\be \label{espace6}
\psi_{j,m}(\vp,x^0) = \sqrt \pi \, f_{j,m}(\tl) \,  e^{2m x^0 +
2 (1+j) \vp} {1
\over \left((2j)!\right)^2}\, .
\ee
In the language of string field theory, this corresponds to
\be \label{espace7}
|\wc\Psi_c^{(2)}\ra = \sum_{j,m} \, {\sqrt \pi\over \left((2j)!\right)^2}  
\,  f_{j,m}(\tl) \, \wh \PP_{j,m} \,
e^{2 m X^0(0)}|0\ra_{X^0} \otimes V_{2 (1+j)}(0)|0\ra_L \otimes
c_1 \, \bar c_1 |0\ra_{ghost}\, .
\ee
The states appearing in \refb{espace7} are precisely the discrete states
of two dimensional string theory\cite{LIAN,9108004} (after inverse Wick 
rotation $X\to i
X^0$.) 

Let us now turn to analyzing the contribution from the rest of the terms 
in $|\BB_2\ra$. From the general formula \refb{es12.4b} we see that this 
will correspond to linear combination of states created from $\wh\PP_{j,m}
\, c_1 \, \bar c_1 e^{2m X^0(0)} V_{Q+iP}(0)|0\ra$ by the action of ghost 
oscillators and the Virasoro generators of the $X^0$ and the Liouville 
field theory. Furthermore by following the same argument as in the case of 
critical string theory we can show that this field configuration must be 
on-shell, {\it i.e.} annihilated by $(Q_B+\bar Q_B)$. The BRST 
cohomology analysis of 
\cite{LIAN} then tells us that these states must be BRST trivial, since 
the only non-trivial elements of the BRST cohomology in the ghost number 
two sector are obtained by taking products of primary states in the matter 
and the Liouville sector with the ground state $c_1\bar c_1|0\ra$ of the 
ghost sector. Thus the part of $|\Psi^{(2)}_c\ra$ other than the one given 
in \refb{espace7} can be removed by a gauge transformation 
$\delta|\Psi_c\ra = (Q_B+\bar Q_B)|\Lambda\ra$. Furthermore, since $Q_B$ 
and $\bar Q_B$ do not mix levels, the gauge transformations which remove 
higher level states built on matter and Liouville primaries cannot affect 
the part of the string field given in \refb{espace7}. Thus we conclude 
that up to a gauge transformation, the effect of $|\BB_2\ra$ is to produce 
the on-shell string field configuration given in 
\refb{espace7}.\footnote{Since earlier we had argued that all 
contributions to $|\Psi^{(1)}_c\ra$ other than the one due to the closed 
string tachyon field can be removed by a gauge transformation, one might 
wonder why we cannot also remove $|\wt\Psi^{(2)}_c\ra$ given in 
\refb{espace7} by a gauge transformation. The reason for this is that the 
states with exponential time dependence were not included in the BRST 
cohomology analysis which led to the conclusion that all contribution to 
$|\Psi^{(1)}_c\ra$ other than the ones coming from the tachyon are BRST 
trivial. This assumption was justified for $|\Psi^{(1)}_c\ra$ which did 
not have any exponential time dependence, but is clearly not justified for 
$|\wt\Psi^{(2)}_c\ra$.}

\sectiono{Matrix Model Description of Two Dimensional String Theory} 
\label{smatrix}

The two dimensional string theory described above also has an alternative
description to all orders in perturbation theory as a matrix
model\cite{GROMIL,BKZ,GINZIN}. This matrix description, in turn, can be
shown to be equivalent to a theory of infinite number of non-interacting
fermions,
each moving in an
inverted harmonic oscillator potential with hamiltonian
\be \label{e12.1}
h(p,q) = {1\over 2} (p^2 - q^2) + {1\over g_s}\, ,
\ee
where $(q,p)$ denote a
canonically conjugate pair of variables.  The coordinate
variable $q$ is related to the eigenvalue of an infinite dimensional
matrix, but this information will not be necessary for our discussion.
Clearly $h(p,q)$ has a continuous energy spectrum spanning the range
$(-\infty, \infty)$.
The
vacuum of the theory corresponds to all states with negative $h$
eigenvalue
being filled and all states with positive $h$ eigenvalue being empty.
Thus
the fermi surface is the surface of zero energy.
Since we shall not go beyond perturbation theory, we shall ignore the
effect of tunneling from one side of the barrier to the other side and
work on only one side of the barrier. For
definiteness we shall choose this to be the negative $q$ side.
In the semi-classical
limit, in which we represent a quantum state by an area element of size
$\hbar$ in the phase space spanned by $p$ and $q$, we can restrict
ourselves to the negative $q$ region, and represent the
vacuum by having the region $(p^2 - q^2) \le -{2\over g_s}$
filled, and rest of the region empty\cite{POLCH,9212027}.
Thus in this picture the fermi
surface in the phase space
corresponds to the curve:
\be \label{e12.2}
{1\over 2} (p^2 - q^2) + {1\over g_s} = 0\, .
\ee

If $\Psi(q,t)$ denotes the second quantized fermion field describing the
above non-relativistic system, then
the massless `tachyon' field in the closed string sector is identified
with the scalar field obtained by the bosonization of the fermion field
$\Psi$\cite{DASJEV,SENWAD,GROSSKLEB}. The precise correspondence goes as
follows.
The classical equation of motion satisfied by the field $\Psi(q,x^0)$ has
the form:
\be \label{e12.1b}
i {\p\Psi\over \p x^0} + {1\over 2} \, {\p^2 \Psi\over \p q^2} + {1\over
2} \, q^2 \Psi - {1\over g_s} \, \Psi=
0\, .
\ee
We now define the
`time of flight' variable $\tau$ that is related to $q$ via the relation:
\be \label{e12.1a}
q = -\sqrt{2\over g_s} \, \cosh\tau\, , \qquad \tau < 0\, .
\ee
$|\tau|$ measures the time taken by a zero energy classical particle 
moving
under the Hamiltonian \refb{e12.1} to travel from $-\sqrt{2\over g_s}$
to $q$. We also define
\be \label{e12.1c}
v(q) = -\sqrt{q^2-{2\over g_s}} = \sqrt{2\over g_s} \, \sinh \tau\, .
\ee
$|v(q)|$ gives the classical velocity of a zero energy particle when 
it is at position
$q$.
Using these variables, it is easy to see that for large negative $\tau$ 
the
solution to eq.\refb{e12.1b} takes the form:
\be \label{e12.1d}
\Psi(q,x^0) = {1\over \sqrt{-2 v(q)}} \, \left[ e^{-i\int^q v(q') dq' +
i\pi/4} \, \Psi_R(\tau, x^0) + e^{i\int^q v(q') dq' -
i\pi/4} \, \Psi_L(\tau, x^0)\right]\, ,
\ee
where $\Psi_L$ and $\Psi_R$ satisfy the field equations:
\be \label{e12.1e}
(\p_0 - \p_\tau) \, \Psi_L(\tau, x^0)=0, \qquad (\p_0 + \p_\tau) \,
\Psi_R(\tau, x^0)=0 \, .
\ee
Thus at large negative $\tau$ we can regard the system as a theory of a 
pair of
chiral fermions, one left-moving and the other right-moving. Of course
there is an effective boundary condition at $\tau=0$ which relates the two
fermion fields, since a particle coming in from $\tau=-\infty$ will be
reflected from $\tau=0$ and will go back to $\tau=-\infty$. Since $\tau$
ranges from 0 to $-\infty$, we can interprete $\Psi_R$ as the incoming 
wave
and $\Psi_L$ as the outgoing wave.

Since $\Psi_L$ and $\Psi_R$ represent a pair of relativistic fermions, we
can bosonize them into a pair of chiral bosons
$\chi_L$ and $\chi_R$. This pair of chiral bosons may in turn be combined
into a full scalar field $\chi(\tau, x^0)$ which satisfy the free field
equation of motion for large $\tau$ but has a complicated boundary
condition at $\tau=0$. If $\chi$ is defined with the standard
normalization, then for large $\tau$ a single right moving fermion is
represented by the configuration\cite{COLEMAN}\footnote{Such a 
configuration has infinite energy at the classical level in the scalar 
field theory. In the fermionic description this infinite energy is the 
result of infinite quantum uncertainty in momentum for a sharply localized 
particle in the position space. Thus the classical limit of the fermionic 
theory does not have this infinite energy. This is the origin of the 
apparent discrepancy between the classical open string calculation of the 
D0-brane 
energy which gives a finite answer and the classical closed string 
calculation which gives infinite answer\cite{0305159}.} 
\be \label{ebo1}
\chi=\sqrt{\pi}\, H(x^0-\tau)\, ,
\ee
and
a
single
left-moving fermion is represented by the configuration
\be \label{ebo2}
\chi=\sqrt{\pi}\, H(x^0+\tau)\, ,
\ee
where $H$ denotes the step
function:
\be \label{es9.7}
H(u) = \cases{1 \quad \hbox{for} \quad u>0\cr
0  \quad \hbox{for} \quad u<0} \, .
\ee

The field $\chi(\tau,x^0)$ is related to the tachyon field $\phi(\vp,
x^0)$ in the continuum description of string theory by a non-local field
redefinition. This relation is easy to write down in the momentum space.
If $\chi(p, x^0)$ denotes the Fourier transform of $\chi$ with respect to
the variables $\tau$, with $p$ denoting the momentum variable
conjugate to $\tau$, and $\phi(P, x^0)$ denotes the Fourier transform
of $\phi$ with respect to the Liouville coordinate $\vp$, with $P$
denoting the momentum variable conjugate to $\vp$, then we
have\cite{9108019,9402156}: 
\be \label{e12.1f}
\chi(P, x^0) = {\Gamma(iP) \over \Gamma(-iP)} \, 
\phi(P, x^0)\, .
\ee
As a result, the background $\chi$ corresponding to the $\phi$ field
configuration given in \refb{e11.7} is given by:
\be \label{e12.1g}
\chi(P, x^0) = -i\, {\sqrt\pi\over
P} \,
\left[ e^{-i P (x^0 + \ln
\sin(\pi\tl)) } +  e^{i P (x^0 + \ln
\sin(\pi\tl)) } \right] \, .
\ee
In $\tau$ space this corresponds to the background:
\be \label{e12.1h}
\chi(\tau, x^0) = -\sqrt{\pi} [H(x^0+ \ln
\sin(\pi\tl) - \tau) -
H(x^0+ \ln
\sin(\pi\tl)+\tau)] + \hbox{constant} \, .
\ee
Eq.\refb{e12.1h} is valid only in the $x^0\to\infty$ limit. Since 
$\tau<0$, in this limit
the first term goes to a constant which can be removed by a redefinition
of $\chi$ by a constant shift, and we get
\be \label{e12.1i}
\chi(\tau, x^0) = \sqrt{\pi} \, H(x^0 + \tau + \ln
\sin(\pi\tl))\, .
\ee
According to \refb{ebo2} this precisely represents a single left-moving 
(outgoing)
fermion. This shows that
the
non-BPS D0-brane of the two dimensional string theory can be identified as
a state of the matrix theory where a single fermion is excited from the
fermi level to some energy
$>0$\cite{0304224,0305194,0305159}.

Since the fermions are non-interacting, the states with a single excited 
fermion do
not mix with any other states in the theory (say with states where two or
more fermions are excited above the fermi level or states where a fermion
is excited from below the fermi level to the fermi level). As a result,
the
quantum states of a D0-brane are in one to one correspondence with the
quantum states of the single particle Hamiltonian $h(p,q)$ given in 
\refb{e12.1} with one
additional constraint, -- the spectrum is cut off sharply for energy below
zero due to Pauli exclusion principle.
Thus in the matrix model description, the quantum `open string field
theory'
for a single
D0-brane is described by the
inverted harmonic oscillator hamiltonian \refb{e12.1} with all the
negative
energy states removed by hand\cite{0308068}. The
classical limit of this quantum
Hamiltonian is described by the classical Hamiltonian \refb{e12.1}, with a
sharp
cut-off on the phase space variables:
\be \label{e12.4}
{1\over 2} (p^2 - q^2) + {1\over g_s} \ge 0\, .
\ee
This is the matrix model description of classical `open string field
theory' describing the dynamics of
a D0-brane. In this description the D0-brane with the tachyon field 
sitting at the maximum of the potential corresponds to the configuration 
$p=0$, $q=0$. The mass of the D0-brane is then given by $h(0,0)=1/g_s$.

Clearly the quantum system described above provides us with a complete
description of the dynamics of a single D0-brane.  In particular
there is no need to couple this system explicitly to closed strings,
although at late time closed strings provide an alternative description of
the D0-brane as a kink solution (given in \refb{e12.1i}). This is in 
accordance with the general
open-closed string duality
conjecture. In the classical limit the 
rolling tachyon solution
in open string theory, characterized by the parameter $\tl$,
corresponds to the phase space trajectory
\be \label{etrajec}
{1\over 2} (p^2 - q^2) + {1\over g_s} = {1\over g_s} \cos^2(\pi\tl)\, ,
\ee
as can be seen by comparing the energies of the rolling tachyon 
system\cite{0203265} and
the system described by the Hamiltonian \refb{e12.1}. In particular the
$\tl\to{1\over 2}$ limit corresponds to a trajectory at the fermi level.

{}From this discussion it is clear
that it is a wrong notion to think
in terms of {\it backreaction of closed string fields on the open string
dynamics}. Instead we should regard the closed string background produced
by the D-brane as a way of characterizing the open string background
(although the open string theory itself is sufficient for this
purpose).
For
example, in the present context, we can think of the closed string tachyon
field $\chi(\tau, x^0)$ at late time as the expectation value of the
operator\footnote{$\p_\tau\hat\chi$ is the representation of the usual
density operator of free fermions in the Hilbert space of first quantized
theory of a single fermion.}
\be \label{eexpec}
\hat \chi(\tau, x^0) \equiv \sqrt{\pi} \, H \left(-\hat q(x^0) -
\sqrt{2\over g_s}
\cosh\tau\right)  \, 
\ee
in the quantum open string theory on a single D0-brane, as described by
\refb{e12.1}, \refb{e12.4}.
In \refb{eexpec} $\hat q$ denotes the position operator in the quantum
open string theory. When we calculate the expectation value of $\hat \chi$
in the
quantum state whose classical limit is described by the trajectory
\refb{etrajec}, we can replace $\wh q$ by its classical value
$q=-\sqrt{2\over g_s} \sin(\pi\tl) \cosh(x^0)$. This gives
\be \label{eexpec2}
\la \hat \chi(\tau, x^0) \ra = \sqrt{\pi} \, H \left(\sqrt{2\over
g_s}
\, \sin(\pi\tl) \, \cosh(x^0) - \sqrt{2\over g_s}
\cosh\tau\right) \simeq \sqrt{\pi} \, H(x^0 + \tau + \ln
\sin(\pi\tl))\, ,
\ee
for large $x^0$ and negative $\tau$. This precisely reproduces 
\refb{e12.1i}.

Given the interpretation \refb{eexpec2} of the closed string field
$|\Psi^{(1)}_c\ra$ produced by the $|\BB_1\ra$ component of the boundary 
state, we
can now ask if it is possible to find
similar interpretation for the exponentially growing component
$|\Psi^{(2)}_c\ra$ of the
string field produced by $|\BB_2\ra$.
We can begin with the simpler task of trying to identify the conserved 
charges $e^{-2m x^0} g_{j,m}(x^0)$ associated with $|\BB_2\ra$. An 
infinite set of conserved charges of this type do indeed exist in the 
quantum theory of a single fermion described by \refb{e12.1}.
These are of the 
form\cite{SENWAD,MOORESEI,UTTG-16-91,9108004,9110021,9209036,9302106,9507041}:
\be \label{econs1old}
e^{-(k-l) x^0} (p+q)^k (q-p)^l\, ,
\ee
where $k$ and $l$ are integers. 
Requiring that the canonical transformations generated by these charges 
preserve the constraint \refb{e12.4} \cite{9108004} gives us a more 
restricted class of 
charges:
\be \label{econs1}
h(p,q) \, e^{-(k-l)
x^0}
(p+q)^k (q-p)^l = \left( {1\over 2} (p^2-q^2) + {1\over g_s}\right) \, 
e^{-(k-l) x^0} 
(p+q)^k (q-p)^l\, .
\ee
Thus it is natural to identify these with linear combinations of
the charges $e^{-2m 
x^0} g_{j,m}(x^0)$ in the continuum theory. In order to find the precise 
relation between these charges we can first compare the 
explicit $x^0$ dependence of the two sets of charges. This gives:
\be \label{econs2}
k-l = 2m\, .
\ee
Thus the conserved charge $e^{-2m
x^0} g_{j,m}(x^0)$ should correspond to some specific linear combination 
of the charges given in \refb{econs1} subject to the condition 
\refb{econs2}:\footnote{Ref.\cite{0307195} 
proposed an alternative route to relating the parameter $\tl$ labelling 
the rolling tachyon boundary state to the parameter labelling the matrix 
model solutions through the ground ring generators\cite{9108004}. However, 
since the ground ring generators are operators of ghost number zero, their 
expectation value on the disk vanishes by ghost charge conservation, and 
hence they have vanishing inner product with the boundary state. Due to 
this reason the relationship between the analysis of \cite{0307195} and 
that given here is not quite clear.}
\be \label{econs3}
g_{j,m}(x^0) \leftrightarrow g_s\, \left( {1\over 2} (p^2-q^2) + 
{1\over g_s}\right) \, \sum_{k\in Z \atop k\ge 0, 2m}\, \left({2\over 
g_s}\right)^{m-k} 
\,  a^{(j,m)}_k \, (p+q)^k 
(q-p)^{k-2m} 
\, .
\ee
Here $a^{(j,m)}_k$ are constants and the various $g_s$ dependent 
normalization factors have been introduced for later convenience.
In order to find the precise form of the coefficients $a_k^{(j,m)}$ we 
compare the 
$\tl$ dependence of the two sides for the classical trajectory 
\refb{etrajec}. Since for this trajectory
\be \label{econs4}
q\pm p = - \sqrt{2\over g_s} \, \sin(\pi\tl) \, e^{\pm x^0}\, ,
\ee
and $g_{j,m}(x^0) = e^{2m x^0} \, f_{j,m}(\tl)$, we have:
\be \label{econs5}
f_{j,m}(\tl) = (-1)^{2m} \, \left(1-\sin^2(\pi\tl)\right) \, 
\sum_{k\in Z \atop k\ge 0, 
2m}\,  
a^{(j,m)}_k \,  \sin^{2k - 
2m}(\pi\tl) \, .
\ee
Thus by expanding $f_{j,m}(\tl)$ given in \refb{efjm} in powers of 
$\sin(\pi\tl)$ we can determine the coefficients $a^{(j,m)}_k$. One 
consistency 
check for this procedure is that on the right hand side the expansion in 
powers of $\sin(\pi\tl)$ starts at order $\sin^{2|m|}(\pi\tl)$. It can be 
verified that the expansion of $f_{j,m}(\tl)$ also starts at the same 
order. The other consistency check is that the right hand side of 
\refb{econs5} vanishes at $\tl={1\over 2}$, which is also the case for 
$f_{j,m}(\tl)$. 

For the purpose of illustration we quote here the non-zero 
values of $a^{(1,0)}_k$ and $a^{(3/2,1/2)}_k$ using \refb{especial}:
\be \label{esp2}
a^{(1,0)}_0 = -2, \qquad a^{(3/2,1/2)}_1 = -3\, .
\ee
In general it follows from the definition \refb{efjm} of $f_{j,m}(\tl)$ 
and the properties of $D^j_{m,-m}(\theta)$ that as a power series 
expansion in $\sin(\pi\tl)$ the maximum power of $\sin(\pi\tl)$ that can 
appear in the expression for $f_{j,m}(\tl)$ is $2j$. Thus the sum over $k$ 
in \refb{econs5} must be restricted to
\be \label{ekjm}
k\le j+m-1\, .
\ee
In other words, $a^{(j,m)}_k$ vanishes for $k>j+m-1$.
We shall now 
show using this fact that the relations \refb{econs3} are invertible, {\it 
i.e.} we can solve them to express $h(p,q) (p+q)^k (q-p)^{k-2m}$ in terms 
of the $g_{j,m}$'s. For this let us restrict to the case $m\ge 0$; the 
$m<0$ case may be analysed in a similar fashion. In this case the sum over 
$k$ in \refb{econs3} runs from $2m$ to $j+m-1$. Thus for $j=m+1$, the only 
term in the sum is $k=2m$. This gives:
\be \label{ekjm1}
g_{m+1,m}(x^0) = g_s \, \left({g_s\over 2}\right)^m \, a^{(m+1,m)}_{2m} \, 
h(p,q) \, (q+p)^{2m}\, .
\ee
This expresses $h(p,q) \, (q+p)^{2m}$ in terms of $g_{m+1,m}(x^0)$. Now 
taking $j=m+2$ in \refb{econs3}, we can express $g_{m+2,m}(x^0)$ as a 
linear 
combination 
of $h(p,q) \, (q+p)^{2m}$ and $h(p,q) \, (q+p)^{2m+1} \, (q-p)$. From
this and \refb{ekjm1} we get $h(p,q) \, (q+p)^{2m+1} \, (q-p)$ in 
terms of $g_{m+1,m}(x^0)$ 
and $g_{m+2,m}(x^0)$. Repeating this process we see that in general 
$h(p,q)\, (p+q)^{2m+l}\, (q-p)^l$ may be expressed as a linear combination 
of $g_{j,m}(x^0)$ for $m+1\le j\le m+l+1$. This shows that the conserved 
charges $g_{j,m}(x^0)$ in the continuum theory contains information about 
the complete set of symmetry generators in the matrix model description of 
the D0-brane.

Since the fields produced by these sources are also proportional to $e^{2m 
x^0} \, f_{j,m}(\tl)$, we expect that the matrix model 
representation of these fields
also involve the same combination as \refb{econs3}. For example we can 
consider the operator:
\ben \label{eoper1}
\hat \chi_{j,m}(\tau, x^0) &\sim&  g_s\, H \left(-\hat q(x^0) -
\sqrt{2\over g_s}
\cosh\tau\right)  \, \left( {1\over 2} 
(\hat p(x^0)^2-\hat q(x^0)^2) +
{1\over g_s}\right) \nonumber \\
&& \, \sum_{k\ge 0, 2m}^{j+m-1}\,  
\left({2\over
g_s}\right)^{m-k}
\,
a^{(j,m)}_k \, 
(\hat p(x^0)+ \hat q(x^0))^k
(\hat q(x^0)-\hat p(x^0))^{k-2m}
\, . \nonumber \\
\een
In this case at late time $\la \p_\tau \, \hat \chi_{j,m}(\tau, x^0) \ra$ 
for the 
classical trajectory \refb{etrajec} behaves as
\be \label{eoper2}
f_{j,m}(\tl) \, e^{2m x^0} \delta(x^0+\tau+\ln\sin(\pi\tl))\, .
\ee
This of course has the same $\tl$ and $x^0$ dependence as 
$\psi_{j,m}(\vp,x^0)$ given in \refb{espace6}, but it is localized in 
$\tau$ 
space at $-(x^0+\ln\sin(\pi\tl))$ by the $\delta$ function instead of 
having the exponential tail of $\psi_{j,m}(\vp,x^0)$ in the negative 
$\vp$ 
region. This however is not necessarily a contradiction, since the fields 
$\la \hat \chi_{j,m}(\tau, x^0) \ra$ are likely to be related to the 
fields $\psi_{j,m}(\vp,x^0)$ by a non-local transformation similar to 
the 
one given in \refb{e12.1f}. Such non-local transformations are known to 
produce exponential tails in the fields in the position space 
representation\cite{9402156}.

In the context we note that by carefully examining the bosonization rules 
for a fermion moving under the influence of inverted harmonic oscillator 
potential, ref.\cite{0401067} has recently argued that in order to 
describe the 
motion of a single fermion in the language of closed string theory, we 
need to switch on infinite number of closed string fields besides the 
tachyon. We believe that the presence of the additional closed string 
background \refb{espace6}, \refb{espace7} associated with the discrete 
states is a reflection of this effect. As a consistency check we note that 
at $\tl=1/2$ the additional background \refb{espace7} vanish. This is 
expected to be true in the matrix model description as well\cite{0401067}.

\end{document}